# Optical Fourier Surfaces for Free-Space Computer-Generated Holography

*Yannik M. Glauser,*[1] *Che-Yung Shen,*[2,3,4] *David B. Seda,*[1] *Çağatay Işil,*[2,3,4]
*Patrick Benito Eberhard,*[1] *Hannah Niese,*[1] *Juri G. Crimmann,*[1] *Daniel Petter,*[1] *Gabriel Nagamine,*[1]
*Sander J. W. Vonk,*[1] *Nolan Lassaline,*[1,6] *Aydogan Ozcan,*[2,3,4,5] *and David J. Norris*[1,*]

[1]Optical Materials Engineering Laboratory, Department of Mechanical and Process Engineering, ETH Zurich, 8092 Zurich, Switzerland

[2]Department of Electrical and Computer Engineering, University of California, 90095 Los Angeles, USA

[3]Department of Bioengineering, University of California, 90095 Los Angeles, USA

[4]California NanoSystems Institute (CNSI), University of California, 90095 Los Angeles, USA

[5]Department of Surgery, David Geffen School of Medicine, University of California, 90095 Los Angeles, USA

[6]Department of Physics, Technical University of Denmark, 2800 Kongens Lyngby, Denmark

*Email: dnorris@ethz.ch

**ABSTRACT.** Computer-generated holography can shape electromagnetic fields by encoding wavefront information in nanostructured surfaces. However, despite significant progress, hologram designs for visible and near-infrared wavelengths remain largely limited by fabrication constraints. Most implementations rely on lithographic methods that produce discretized surfaces that do not match the wave nature of light. Further advances would benefit from accurate continuous (grayscale) control of interfacial profiles, which would allow straightforward design principles from Fourier optics to be applied. In this work, we introduce optical Fourier surfaces as a versatile, intuitive platform for free-space computer-generated holograms. Exploiting these arbitrarily wavy surfaces, we demonstrate three different design strategies for computer-generated holography. First, we use an analytical linear design approach based on sinusoidal lenses and gratings as fundamental holographic building blocks. Second, we enhance diffraction efficiencies through an iterative Fourier-transform algorithm. Finally, we extend our framework with machine-learning-based inverse design, creating wavelength-multiplexed holograms that reconstruct distinct diffraction responses in a predefined image plane. Our results combine advanced thermal scanning-probe lithography with flexible design strategies to create useful structures for photonic applications, particularly in settings where rapid prototyping is crucial.

## INTRODUCTION

The ability to control light is key for many technologies, including imaging,[1,2] augmented reality,[3] communication,[4,5] and computing.[6,7] For such applications, holography presents a general framework for tailoring the optical wavefront in terms of its amplitude, phase, polarization, and spectrum. In its traditional form, holography involves a two-step process:[8,9] (i) an interference pattern between an object and reference wave is recorded as a hologram in a medium, and (ii) this hologram is later re-illuminated with the reference wave to generate an image of the original object. However, the need for a physical recording step can be limiting for applications. To avoid this, computer-generated holography (CGH) has been developed.[9] It calculates the required hologram directly using computer algorithms, offering a route to full digital control over the information encoded in the electromagnetic field.

To implement this approach in practice, the hologram must then be fabricated from the computer-generated design. For visible and near-infrared light, the desired structures often contain surface profiles that continuously vary in depth ($z$).[10-17] Unfortunately, most methods that tailor interfacial depth (known as grayscale lithography) have not yet demonstrated sufficient control for the required "wavy" profiles. Consequently, researchers have restricted their designs to surface profiles with only two depth levels.[18] While, in general, this constraint reduces optical performance, properly arranged arrays of "binary" nanostructures, known as metasurfaces, can collectively provide advanced capabilities.[19,20] These structures offer independent control over amplitude,[21-25] phase,[19,22-26] polarization,[21,24-27] and/or spectrum[28] by engineering subwavelength building blocks. However, the design process for metasurfaces is usually highly specialized and non-intuitive, relying on computationally intensive full-field simulations to achieve a design with the desired optical response. These difficulties arise because metasurfaces force binary elements to shape electromagnetic waves—a task that is presumably better matched to wavy patterns. This suggests that more effort should focus on improved grayscale solutions for CGH that combine highly accurate nanofabrication with quick and intuitive design.

Recently, thermal scanning-probe lithography (TSPL) has been introduced as a powerful method to obtain grayscale surface patterns with nanometer-scale accuracy.[29-32] Because this technique enables surface profiles with precise depth control, it can provide arbitrarily wavy diffractive interfaces, known as optical Fourier surfaces (OFSs).[33] To date, OFSs made by TSPL have primarily been explored in two distinct contexts. First, they have served as model systems to perform fundamental investigations of diffraction from high-fidelity sinusoidal gratings.[34] Second, they have been incorporated into architectures that target integrated photonics, e.g., grating couplers, waveguide resonators, local sources/sensors, and optical-computing elements.[35-38] These developments demonstrate the power of OFSs, but leave CGH with freely propagating waves unexplored. Owing to their direct connection to Fourier optics, OFSs provide a natural platform for realizing compact, high-fidelity holograms in the visible and near infrared.

Here, we use TSPL to fabricate wavy diffractive surfaces in silver (Ag) for CGH. We employ three complementary design strategies: (i) analytical linear design, (ii) iterative design using a generalized Gerchberg–Saxton (GS) algorithm,[39,40] and (iii) machine-learning-based inverse design.[6,41,42] While each approach offers distinct strengths and limitations, all of them can be seamlessly implemented with our OFSs. By combining these flexible design strategies with our accurate fabrication procedure, we realize compact, high-fidelity OFS holograms capable of producing sophisticated diffraction patterns.

## RESULTS AND DISCUSSION

**Optical Fourier surfaces as reflective phase holograms.** We fabricated Ag OFSs by applying a previously reported procedure.[35] Briefly, we performed the following steps: (i) thermally sensitive poly(phthalaldehyde) (PPA) thin films were patterned with TSPL,[29-31] (ii) Ag was then thermally evaporated onto the PPA,[43] (iii) the Ag replica was template-stripped off of the PPA,[44] and (iv) the final Ag structure was cleaned with anisole (see Methods and Figure S1 in the Supporting Information). Compared to our prior work, we achieved 20% deeper structures (with depths of up to 350 nm) by improving the feedback calibration of the TSPL tool.

The resulting Ag OFSs act as phase-only reflective holograms by locally introducing optical-path differences to the incoming reference wavefront with free-space wavevector $k_0$ (Figure 1a). The height profile $h_\mathrm{p}(x, y)$ imprints a phase modulation $\phi(x, y) = 2k_0 h_\mathrm{p}(x, y)$ at the sample plane. After free-space propagation of the reflected light, a complex diffraction pattern $U(x, y)$ appears at a target image plane.[9,34,45,46] In this context, CGH using OFSs involves the direct design of the phase modulation introduced by the surface profile. Even though the surface profile only alters the phase in the sample plane, manipulation of the amplitude and/or phase of the light in the far field (Fourier space) or at any arbitrary image plane is possible.[35] However, in free space, inherent trade-offs exist between the image quality and diffraction efficiency. The spatial resolution of OFSs (down to ~30 nm in-plane and ~2 nm in depth for shallow structures) allows these trade-offs to be optimized.

Prior works have developed strategies to address CGH for surfaces that only affect the phase. Typically, these approaches exploit Fourier optics and scalar diffraction theory.[9,47] In the most general case, the desired image is formed in an arbitrary image plane located at a specific distance from the diffractive hologram. Light propagation is analytically described using the angular-spectrum method.[9,48] The two-dimensional wavefront at the sample plane (input plane in $x$ and $y$) is decomposed into its spatial Fourier components, which then propagate in the $z$ direction according to a transfer function. These propagated Fourier components are subsequently recombined to reconstruct the diffraction signal in the image plane (output plane). A special case arises when the diffraction image is formed in the far field. Light propagation to the far field (Fourier space) is then simply described by a Fourier transform of the electric field profile in the sample plane.[9] This can be realized in an optical setup by imaging the back focal plane of a $2f$ imaging system, where $f$ is the focal length. (See Methods and Figure S2 for experimental details and Sections S1.1 and S1.2 in the Supporting Information for diffraction theory.)

**Analytical linear design.** To design OFSs for CGH, we must determine the surface profile (i.e., the phase modulation at the sample plane) that creates the desired diffracted signal at the output plane. For shallow profiles, we can follow a linear approach.[33,35] Namely, we can simply superpose individual diffractive elements to create the desired output. Mathematically, this follows from a first-order Taylor expansion of the phase $e^{i\phi} \approx 1 + i\phi$. This approximation is valid for small phase modulations and inherently neglects higher-order contributions.

For our fundamental diffractive element, we consider sinusoidal lenses (Fresnel zone plates, Figure 1b) or sinusoidal gratings (Figure 1g). Each produces distinct, point-like diffraction features: a tight focus and discrete diffraction spots, respectively (Figure 1b–g). Superposition of these building blocks enables the construction of more advanced diffracted images. In contrast, the superposition of other elements, such as blazed Fresnel lenses or blazed gratings, simply results in another lens or grating with only one diffracted feature, offering no additional functionality (Section S1.3 and Figure S3 in the Supporting Information).

TSPL allows us to exploit such wavy sinusoidal lenses (Figure 1d) and sinusoidal gratings (Figure 1f) as building blocks. Even its high lateral spatial resolution, down to ~30 nm for shallow structures, is beneficial. While in principle high spatial frequencies (corresponding to periods, $\Lambda$, shorter than the optical wavelength, $\lambda$) only generate evanescent waves that do not propagate to the far field, they still influence the diffraction output through higher-order convolution effects. We show analytically (Section S1.4 and Figure S4 in the Supporting Information) that all spatial frequencies contribute to the output signal. However, the contributions of structural features with periods $\Lambda < \lambda/4$ are small.

While the linear approach is valid for any image plane, we restrict our discussion now to the far field. In this case, the inverse Fourier transform $\mathcal{F}^{-1}$ of a desired complex-valued diffraction image yields the corresponding surface profile (up to a scaling factor), enabling the direct and intuitive design of OFS holograms (Figure 2a). To ensure that $\mathcal{F}^{-1}$ results in a real-valued phase map that can be introduced by a single reflective surface profile, the tunable part of the desired diffraction pattern $\mathcal{F}\{\phi\} = U(x, y)$ needs to be Hermitian, with $U(-x, -y) = U^*(x, y)$. This requirement arises because only conjugate-symmetric Fourier spectra correspond to real-valued functions in real space (see Section S1.5 in the Supporting Information).

We first investigate OFSs formed by azimuthal superpositions of sinusoidal profiles (see Section S1.6 and Figures S5 and S6 in the Supporting Information). Each sinusoid corresponds to a specific spatial frequency that produces a pair of diffraction spots in the far field ($-1^{st}$ and $+1^{st}$ orders symmetrically arranged around the $0^{th}$-order spot). Therefore, superposing many sinusoids enables us to diffract light simultaneously into various directions, forming a desired pattern in the far field (Fourier space). In particular, holograms with useful diffraction patterns for photonic applications[49] can be generated.

For example, we fabricated an OFS hologram that projects the ETH logo (Figure 2b). To prevent overlap with the specular reflection, the target image is offset in Fourier space. The specular reflection was eliminated using a beam block. The corresponding simulated and experimentally measured far-field intensity patterns are shown in Figure 2c,d. The results are in good agreement with the targeted diffraction image. Analogous results for a grayscale image (a cube with facets of varying intensities) are presented in Figure S7 in the Supporting Information. Note that as more sinusoids are superposed, the contribution from each becomes weaker (Figure S6 in the Supporting Information). Thus, the overall intensity in Figure 2d remains relatively low compared to the specular reflection. The simulated and experimental efficiencies were 0.18 and 0.11% per image (0.36 and 0.21% for both combined), respectively (Figure 2c,d and Figure S8 in the Supporting Information).

Fortunately, in some applications, the overall efficiency is not the most important characteristic. Instead, the image quality in terms of its uniformity and sharpness plays a more critical role.[50,51] For our structures, the obtained images are uniform in intensity, confirming that not only the amplitude but also the phase of the diffracted light is controlled. This is expected from theory, as phase information is preserved in the Fourier-based design (see simulated data in Figure S9 in the Supporting Information). This reduces speckles that cause non-uniform diffraction images, a common problem in CGH.[52]

To enhance the diffraction efficiency of the hologram within the linear approach, we can rescale the surface profile to boost the amplitude of each sinusoidal component. By keeping the total pattern depth fixed, this scaling process leads to cropping of the dominant sinusoidal features in the center of the OFS (Figure S10 in the Supporting Information). As the scaling factor increases, the diffraction efficiency of the hologram improves. However, with increasing pattern depth, the phase modulation progressively departs from the small-phase regime assumed by the linear approach. Effects due to higher-order diffraction become more dominant, resulting in images with reduced uniformity and more pronounced contours.

Although we have so far focused on far-field images, the linear design approach can also be applied to arbitrary image planes. However, in these cases, the fundamental building blocks are sinusoidal lenses, each focusing light to a specific location in the image plane (Figure 1b–d). By superposing these lenses, desired focal patterns can be created.[35]

**Iterative design.** While the linear approach is intuitive and offers control over amplitude and phase, some applications require higher diffraction efficiencies. Deeper profiles can provide this efficiency, but the relationship between the profile and signal is then nonlinear. Namely, the superposition of two patterns with phases $\phi_1(x, y)$ and $\phi_2(x, y)$ in the surface results in a product of complex exponentials $e^{i(\phi_1+\phi_2)} = e^{i\phi_1}e^{i\phi_2}$. In the far field, this causes a convolution of the individual diffraction signals,

$$\mathcal{F}\{e^{i\phi_1+i\phi_2}\} = \mathcal{F}\{e^{i\phi_1}\} * \mathcal{F}\{e^{i\phi_2}\} \neq \mathcal{F}\{e^{i\phi_1}\} + \mathcal{F}\{e^{i\phi_2}\}\,. \quad (1)$$

In other words, the linear superposition of surface profiles does not yield a linear superposition of the corresponding diffraction patterns, but rather their convolution.

For this case, iterative design strategies can be applied.[47] This results in less intuitive designs that impose fewer constraints, e.g., no Hermitian symmetry is required. The most commonly used methods for far-field holography are iterative Fourier-transform algorithms (IFTAs). They are also more broadly exploited in electron microscopy, astronomy, and crystallography.[39,40] Essentially, they allow phase information to be reconstructed from intensity measurements. Consequently, they are often referred to as phase-retrieval algorithms. Variations of IFTAs include steepest-descent, input–output, and error-reduction methods, with the Gerchberg–Saxton (GS) algorithm the most widely used among the latter.[39,40,53]

To complement our linear design approach, we investigated a generalized form of the GS algorithm that iteratively refines the complex wavefront at the object domain (sample plane, real space) and the Fourier domain (image plane, Fourier space). In between, both object- and Fourier-domain constraints are imposed. See Section S1.7 and Figure S11 in the Supporting Information for a detailed description of the procedure.

We employed this iterative approach to project the ETH logo and a grayscale cube. Without intervention, the GS algorithm introduces discontinuities in the surface profile (phase jumps from 0 to $2\pi$), which are not possible to fabricate with TSPL. To remove these, we post-processed the design by applying a Gaussian filter (standard deviation $\sigma$ = 60 nm) to smoothen the surface and generate the final design for fabrication. Figure 3a,b shows the fabricated holograms, which exhibit different structural features compared to the linear designs. Figure 3c,d plots the corresponding far-field diffraction patterns, alongside a simulation based on scalar-diffraction. For the ETH logo, the target images are well reconstructed in both simulation and experiment with efficiencies of 57.6 and 32.9%, respectively. The measured value is significantly higher than that from the linear approach. However, the diffraction images exhibit speckle patterns and reduced uniformity. As discussed above, these artifacts are a consequence of the unconstrained phase in the Fourier domain, which can introduce undesired interference.[52]

To improve uniformity, several variations of the GS algorithm have been reported that constrain the phase of the wavefront in the Fourier domain.[52,54] We tested this strategy by adapting the iterative algorithm for our OFS platform. It produced designs similar to those obtained directly from the linear approach (see Figure S12 in the Supporting Information). In these cases, the final iteration largely dictated the design outcome, as it applied the inverse Fourier transform after enforcing both amplitude and phase constraints in the Fourier domain (analogous to the linear approach). This observation highlights that the surface designs produced by different approaches converge, when similar constraints are applied.

**Machine-learning-based inverse design.** To further explore the capabilities of our OFS holograms, we pursued a stochastic gradient descent-based differentiable inverse-design framework that used deep learning for diffractive feature optimization.[6,41] This enabled access to a broader set of design solutions that overcome the limitations of linear or standard iterative design approaches. We leveraged it to realize wavelength-multiplexed holograms that project diffraction images at arbitrary image planes, moving beyond the conventional far-field regime. More specifically, the design algorithm optimized the surface profile at the sample plane such that, upon free-space propagation, it generated the desired intensity patterns, at specific wavelengths, in a target image plane located at a defined distance from the OFS hologram. During deep learning-based optimization of diffractive features, a mean-squared-error (MSE)-based loss function between the output intensity profiles and their corresponding ground truth images was minimized using error backpropagation and gradient descent.[55] In addition to the image-fidelity term, the loss function incorporated output diffraction efficiency-based constraints and fabrication-aware penalty terms that enforced a maximum allowable profile slope of 1 (45°), thereby reducing potential discrepancies between simulations and experimental results (see Section S1.8 in the Supporting Information). To further improve robustness, a Gaussian filter ($\sigma$ = 320 nm) was also applied to the surface profile during optimization. We assumed plane wave illumination, which is experimentally approximated by a Gaussian beam with a large beam diameter (see Methods and Figure S2b in the Supporting Information).

As a proof of concept, we designed, fabricated, and optically characterized a wavelength-multiplexed hologram that reconstructs distinct images at three different wavelengths ($N_w$ = 3). Figure 4a,b shows the original structure in PPA and the OFS hologram in Ag, respectively. The corresponding target images in Figure 4c,f,i are plotted alongside the simulated images in Figure 4d,g,j, which are calculated based on the angular-spectrum method.[9,48] They represent handwritten digits 0, 1, and 2 from the Modified National Institute of Standards and Technology (MNIST) dataset (each image with 25×25 pixels) assigned to wavelengths $\lambda$ = 450, 575, and 700 nm, respectively, reconstructed at an image plane located 200 μm away from the hologram. Figure 4e,h,k shows the experimentally measured diffraction images for the different wavelengths. The desired images are diffracted into a field of view (FOV) of 50 μm × 50 μm (indicated by the dashed square in Figure 4e) with residual intensity distributed around this region. While good agreement between the target and diffracted images is observed, the experimental results exhibit lower overall intensity and increased crosstalk between the wavelength channels compared to the simulations. Quantitative values for diffraction efficiency and image quality (peak signal-to-noise ratio, PSNR) are provided in Table S1 in the Supporting Information. These discrepancies are likely due to fabrication imperfections, experimental setup-related artifacts and potential misalignments. In addition, we observe speckles within the reconstructed images, which we attribute to the unconstrained phase of the wavefront in the design process.

To gain deeper insight into the behavior of a wavelength-multiplexed hologram, we systematically analyzed several key design parameters: (i) the distance between the OFS hologram and image planes, (ii) the spectral spacing, and (iii) the number of multiplexed channels $N_w$ (see Figures S13 and S14 in the Supporting Information). These studies revealed clear trade-offs in the information content encoded in a single OFS, influenced by, e.g., image complexity, reconstruction quality, crosstalk, diffraction efficiency, robustness, and fabrication feasibility. In particular, increasing the number of wavelength channels places greater demands on the surface profile, making it necessary to reduce the complexity of each target image to maintain reliable reconstruction across all channels. Guided by this analysis, we found that four-wavelength ($N_w$ = 4) spectral multiplexing could be improved by reducing the target diffraction-image resolution from 25 × 25 to 7 × 7 pixels, thereby relaxing the information-density requirements imposed on a single OFS hologram. To experimentally demonstrate this design strategy, we fabricated some representative four-channel OFS holograms. Figure 5a–c shows the patterned structures in PPA for three different image series: geometric shapes (squares and rectangles), the letters "ETHZ," and the letters "UCLA," respectively. The corresponding OFS holograms in Ag are shown in Figure 5d–f. Simulated and experimentally measured diffraction images at an image plane located 200 μm from the sample plane are plotted in Figure 5g–l for wavelengths $\lambda$ = 440, 540, 640, and 740 nm. The measured intensity patterns inside the FOV were binned to match the simulation resolution (see raw experimental data in Figure S15 in the Supporting Information). The corresponding efficiencies and PSNR values are summarized in Table S2 in the Supporting Information. Collectively, these results demonstrate that machine-learning-based inverse design can flexibly balance image complexity, number of spectral channels, diffraction efficiency, and fabrication constraints, providing a general strategy for encoding arbitrary wavelength-multiplexed optical functions into an OFS hologram.

## CONCLUSIONS

In this work, we introduce OFSs as a versatile, previously unexplored platform for free-space CGH in the visible and near-infrared regime. Combining nanometer-scale fabrication through TSPL with analytical and data-driven design approaches, we demonstrated compact holograms that encode diffraction patterns at high spatial resolution. Our improved fabrication process—based on an optimized feedback calibration of the TSPL tool—enables high-fidelity OFSs with increased depth. This advancement supports the production of holograms across a broader design space. We investigated design methodologies with varying levels of complexity and reconstruction performance. Starting with sinusoidal gratings and lenses as intuitive building blocks, we developed a linear superposition framework that provides direct analytical control over amplitude and phase in the far field. To boost diffraction efficiencies, we applied a generalized GS algorithm, which relaxes first-order approximations at the cost of phase control, leading to reduced homogeneity in the reconstructed images. Finally, we employed machine-learning-based inverse design to create wavelength-multiplexed holograms that project distinct patterns at different wavelengths onto a predefined image plane. We experimentally demonstrated up to 4-fold multiplexing. This approach effectively adapts to variations in image resolution, spectral spacing, number of wavelength channels, or distance between sample and image plane. Altogether, the design flexibility and simplicity of OFSs combined with their accurate fabrication procedure provides benefits for applications in rapidly evolving fields where fast prototyping is essential.

## METHODS

**Design of Fourier surfaces and bitmap generation.** Our designs of the OFS holograms were generated either from analytical functions [for gratings, lenses, and (quasi)crystals] or from one of the three design strategies outlined above: linear design, iterative design via IFTAs, or machine-learning-based inverse design. The feasibility of these designs is governed by the limitations of TSPL fabrication, including constraints on maximum lateral dimensions, depth, and gradient. While the lateral OFS dimension is limited by the piezo stage scan range of ~50 µm, our experiments indicate that surface slopes exceeding ~1 (45°) are currently not reproducible. All our design strategies explicitly incorporate these fabrication constraints.

Each hologram design is represented by an 8-bit map. We use pixel sizes down to 20 nm × 20 nm. Before loading the final designs into the TSPL tool, they are inverted along the $z$ axis and mirrored in the $yz$ plane to account for template stripping during fabrication.

**Fabrication procedure.** The detailed fabrication procedure has been reported previously.[35] For TSPL, PPA was used as a thermal resist to reduce tip contamination and increase sample yield compared with other previously used resist materials.[33] In contrast to earlier work,[33,34] we additionally performed a cleaning step in anisole (AR 600-02, Allresist) for 2 min to remove residual PPA on the Ag. This enabled the fabrication of significantly larger and deeper structures without PPA contamination.

We also improved the feedback calibration of the TSPL tool to increase the accessible pattern size and depth. To this end, the feedback was trained by patterning structures with gradually increasing depths until the desired depth, up to 350 nm, was reached. Depending on the complexity of the pattern, the writing time per pixel was increased from the default value of 25 to 50 µs, while the in-situ topography readout was upsampled by a factor of 2.

**Characterization of surface topography.** The accuracy of our fabrication procedure has been demonstrated in previous studies.[33,34,56] To verify the updated process, which includes additional anisole cleaning, we used a reference structure: a single-sinusoidal profile with period $\Lambda$ = 600 nm, amplitude $A$ = 25 nm, and lateral dimensions of 9 µm × 9 µm. The sinusoidal structure is surrounded by a 1-µm-wide patterned frame (see Figure S16 in the Supporting Information).

First, we investigated the pattern transfer from PPA to Ag by comparing the original structure in PPA with the final structure in Ag. Both topographies were measured by atomic force microscopy (AFM). We leveled the measured profiles by performing a two-dimensional planar fit based on the unpatterned areas surrounding the OFS. Afterwards, we fit a sinusoidal function to the patterned surface profile. The characterization was performed using a function-fitting software package (FunFit).[57] The corresponding data (fit parameters) can be found in Table S3 in the Supporting

Information. The results confirm that the surface profile was not changed during pattern transfer to Ag (deviations in periods and amplitudes are ~1 nm). Moreover, we compared the roughness for the patterned and unpatterned areas in Ag. In line with our previous findings,[33] we observed that the patterned areas exhibited smaller surface roughness (1.4 versus 1.9 nm root-mean-square roughness).

Second, we compared the topography data obtained from the TSPL tool with the AFM topography data of the same structure in PPA. We found good agreement between both methods (see fit parameters in Table S3 in the Supporting Information).

We concluded that: (i) the pattern transfer from PPA to Ag does not alter the surface structure, and (ii) the TSPL tool provides topography data comparable to AFM topography measurements. Thus, the TSPL data obtained during patterning of the PPA was sufficient for the surface characterization of the OFS holograms in Ag. This is consistent with our previous work.[34]

**Optical measurements.** For both the far-field diffraction experiments and those with an image plane at an intermediate distance from the sample plane, we used the same optical setup in two slightly different configurations (Figure S2 in the Supporting Information). The home-built setups are based on an inverted microscope body (Eclipse Ti−U, Nikon).

As a coherent light source, we used a supercontinuum laser (SuperK Fianium, NKT Photonics) combined with a tunable band-pass filter (wavelength range 400–1000 nm, linewidth ~1.5 nm; LLTF Contrast, NKT Photonics). While the wavelength for the far-field experiments was fixed to 532 nm, the wavelength was adjusted for the spectral-multiplexing experiments. The laser beam was collimated with a 10× objective [TU Plan Fluor, numerical aperture (NA) of 0.3; labeled L1 in Figure S2], after being launched from an optical fiber (A502-010-110, NKT Photonics). The light then passed through a 750-nm short-pass filter (FESH0750, Thorlabs; SP) to reject residual near-infrared light coupled out by the tunable band-pass filter. Next, the light was passed through a 90:10 beamsplitter (BSN10R, Thorlabs; BS1), a linear polarizer (WP25M-VIS, Thorlabs; P1), several mirrors, a defocusing lens (AC254-400-A-ML, Thorlabs; L2), and a 50:50 beamsplitter (AHF analysentechnik; BS2) before being focused onto the back focal plane of a 50× objective (TU Plan Fluor, NA of 0.8; L3) to achieve wide-field illumination. This resulted in a broad Gaussian beam at normal incidence (best experimental approximation of a plane wave), which was reflected from the OFS hologram. A power meter (PM100D, Thorlabs; PM) was used to monitor laser-power fluctuations over time.

To measure the far-field response of the holograms, we used a Fourier microscope configuration.[58] While the OFS hologram was positioned in the front focal plane of the objective (real plane), emanating angles were focused to points on the back-focal plane (Fourier plane, representative of the far field). The back-focal plane was relayed to a digital camera (Zyla PLUS sCMOS, Andor) after passing lenses L4–L7 (tube lens, Nikon; AC254-200-A-ML, Thorlabs; AC254-200-A-ML, Thorlabs; AC508-200-A-ML, Thorlabs, respectively), an iris, and another linear polarizer (WP25M-VIS, Thorlabs; P2). To collect only light that interacted with the OFS, we placed the iris in a real plane and adjusted its diameter to match the structure side length.

To measure the diffraction response in an image plane at a specific distance from the sample plane, we defocused the sample by the desired distance and removed the Fourier lens L6. In this configuration, the desired image plane was in focus, representing a real plane in the imaging system which is mapped onto the camera. The iris was removed from the collection path of the setup.

The polarization of the incident and collected light could be independently analyzed with the polarizers (P1, P2). Because plasmonic coupling effects were negligible for our OFS holograms (p-polarized light can couple to surface plasmon polaritons under momentum-matching conditions[33,34,59]), we did not observe significant differences in diffraction response for different input polarizations. Therefore, both polarizers were fixed to pass light polarized along the $x$ axis.

Precise alignment of the optical setup is essential for reliable diffraction measurements. We followed the same alignment procedure described previously.[34] This procedure helped to ensure that the measured signals originated from the OFS holograms rather than from misalignment, thereby improving reproducibility and reducing systematic errors.

**Characterization of diffraction images**. The diffraction efficiencies of the simulated and experimentally measured far-field holograms (data in Figs. 2c,d and 3b) were determined by integrating the intensity counts over all the pixels contributing to the diffraction image. These integrated counts were then divided by the total measured counts to obtain the efficiency (see Figure S8 in the Supporting Information).

The diffraction efficiencies and PSNR values for the wavelength multiplexed holograms (data in Figs. 4c–k and 5g–l) were calculated based on eqs S59 (for $N_w$ = 1) and S61 in the Supporting Information. Because the input intensity distribution on the OFS hologram is difficult to access experimentally, the diffraction efficiency for each wavelength was determined by normalizing the intensity counts within the FOV to the total counts in the image. In this evaluation, absorption losses in Ag (estimated to be a few percent across the considered wavelength range[33]) were not included.

The efficiency and PSNR values are summarized in Tables S1 and S2 in the Supporting Information. The experimentally measured diffraction efficiencies are higher than the simulated values, particularly for wavelengths near the center of the spectral range. This difference likely arises from increased crosstalk between channels in the experiments, which introduces additional signal within the FOV. While this boosts the measured efficiency, it reduces the PSNR.

**ASSOCIATED CONTENT**

**Supporting information.**

Discussion of scalar diffraction theory and paraxial diffraction models (Sections S1.1 and S1.2). Additional information on fundamental diffractive building blocks (Section S1.3) and subwavelength features (Section S1.4). Further explanations on the different design approaches (Sections S1.5–1.8). Additional figures and tables (Sections S2 and S3) discussing the fabrication process, the optical setup, and additional simulated and experimental data.

**AUTHOR INFORMATION**

**Corresponding author**

*Email: dnorris@ethz.ch

**Email and ORCID**

Yannik M. Glauser: yglauser@ethz.ch, 0000-0002-5362-0102

Che-Yung Shen: steven121200@g.ucla.edu, 0009-0003-0546-6344

David B. Seda: daseda@ethz.ch, 0009-0007-4539-169X
Çağatay Işil: cagatayisil@g.ucla.edu, 0000-0003-3367-1858
Patrick Benito Eberhard: peberhard@ethz.ch, 0009-0002-8824-1038
Hannah Niese: nieseh@ethz.ch, 0000-0002-5980-5386
Juri G. Crimmann: jcrimmann@ethz.ch, 0000-0002-0367-5172
Daniel Petter: daniel.petter@papura.at, 0000-0003-2788-1138
Gabriel Nagamine: gabriel.nagaminegomez@unibe.ch, 0000-0002-4830-7357
Sander J. W. Vonk: vonks@ethz.ch, 0000-0002-4650-9473
Nolan Lassaline: nlasso@dtu.dk, 0000-0002-5854-3900
Aydogan Ozcan: ozcan@ucla.edu, 0000-0002-0717-683X
David J. Norris: dnorris@ethz.ch, 0000-0002-3765-0678

**Note**

The authors declare the following competing interests: N.L. and D.J.N. hold a patent related to optical Fourier surfaces.

**ACKNOWLEDGMENTS**

This project was funded by ETH Zurich. S.J.W.V. acknowledges support from the Swiss National Science Foundation (grant 200021-232257). N.L. acknowledges funding from the Swiss National Science Foundation (Postdoc Mobility P500PT_211105) and the Villum Foundation (Villum Experiment 50355). A.O. acknowledges the support of US National Science Foundation. We thank J. J. Erik Maris and A. C. Hernandez Oendra for stimulating discussions and J. Chaaban and U. Drechsler for technical assistance.

**FIGURES**

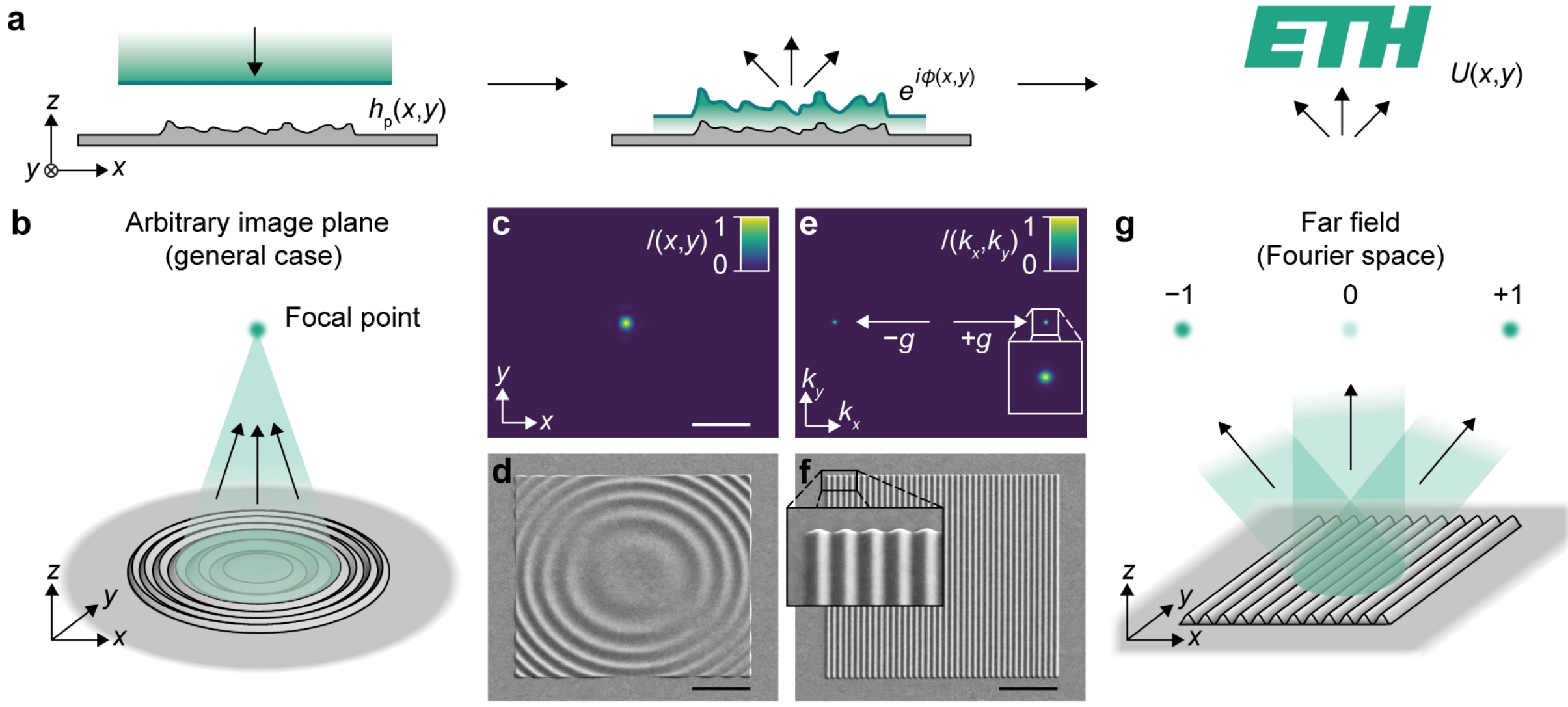

**Figure 1. Optical Fourier surfaces (OFSs) in Ag as reflective phase-only holograms.** (a) Computer-generated holography (CGH) with OFSs. The OFS hologram is designed and fabricated with a height profile $h_{\mathrm{p}}(x, y)$ that induces a phase modulation $\phi(x, y)$ on the reflected and diffracted light in the sample plane, thereby shaping the complex wavefront $U(x, y)$ in an arbitrary image plane. (b) Schematic of a sinusoidal lens with focal distance $f$, focusing light to a specific spot. (c) Experimentally measured focal spot for 532-nm light of a sinusoidal lens in Ag at a distance $f$ = 100 µm. (d) Scanning electron microscopy (SEM) image of the corresponding lens. (e) Experimental far-field diffraction pattern of a sinusoidal grating with a period $\Lambda$ = 1000 nm. (f) SEM image of the corresponding grating in Ag. (g) Schematic of sinusoidal OFSs diffracting light into the far field (Fourier space). SEM images in (d) and (f) were acquired at a tilt angle of 30°. The scale bar is 10 µm. Intensities in (c) and (e) are normalized to their respective maximum values and plotted on a linear scale.

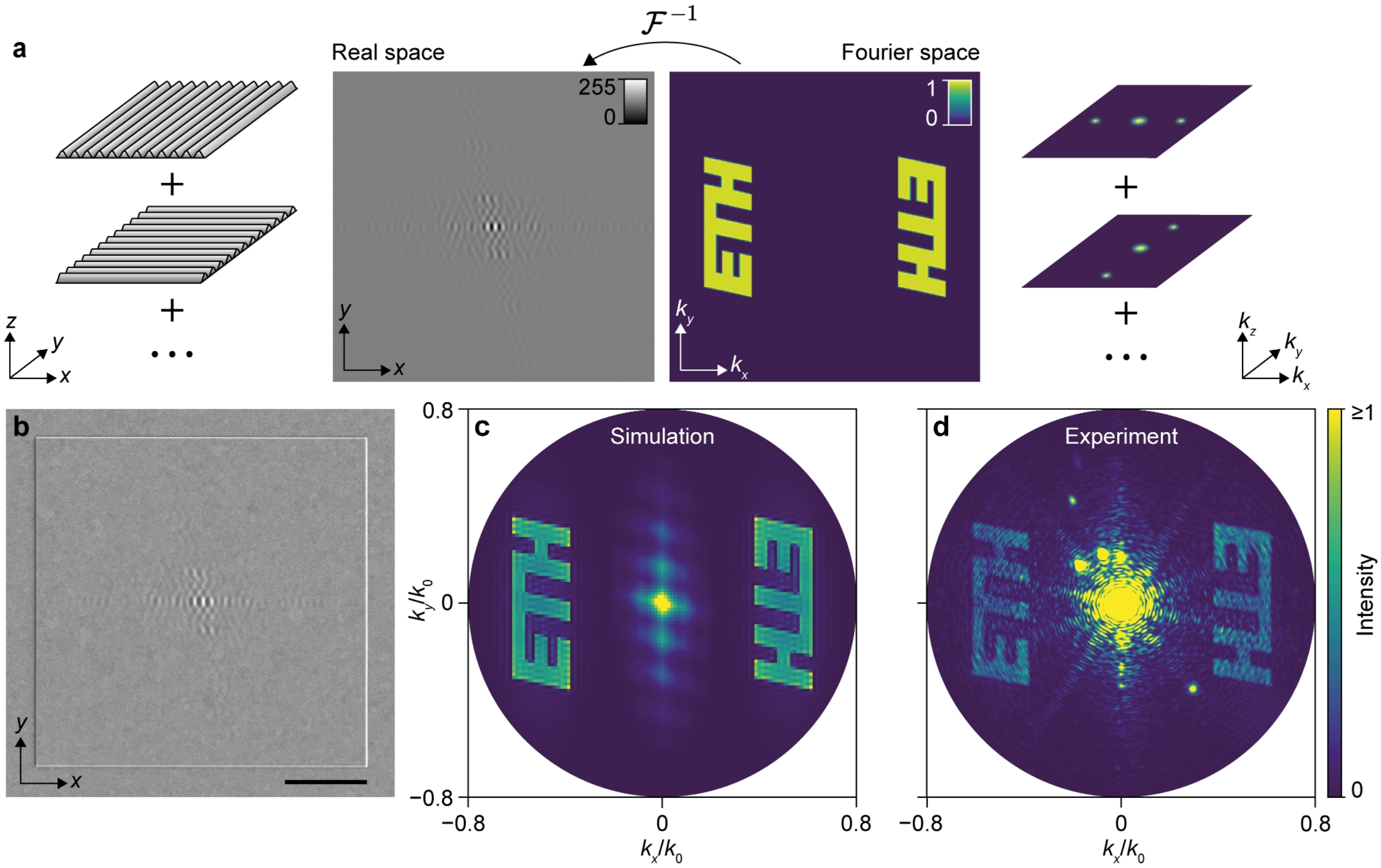

**Figure 2. Linear design approach for far-field holography with OFSs.** (a) Inverse Fourier transform of the target diffraction pattern (complex-valued) in Fourier space to obtain the corresponding OFS design in real space (sample plane). We use a reference wave at normal incidence with $\lambda = 532$ nm. The OFS is rescaled to a grayscale map for TSPL (8-bit precision). (b) SEM image of the fabricated OFS hologram in Ag, designed to produce a diffraction image in the form of the logo of ETH Zurich. During fabrication, the design in (a) was converted to a surface profile with a target depth of $\lambda/2 = 266$ nm. The scale bar is 10 µm. (c) Simulated and (d) experimentally measured far-field diffraction intensities in Fourier space (i.e., the back focal plane of the objective with a numerical aperture, NA of 0.8). Intensities in (c) and (d) are normalized to the respective maximum values of the diffraction image (excluding the $0^{th}$-order specular reflection) and plotted on a linear scale. The simulation is based on scalar diffraction theory.

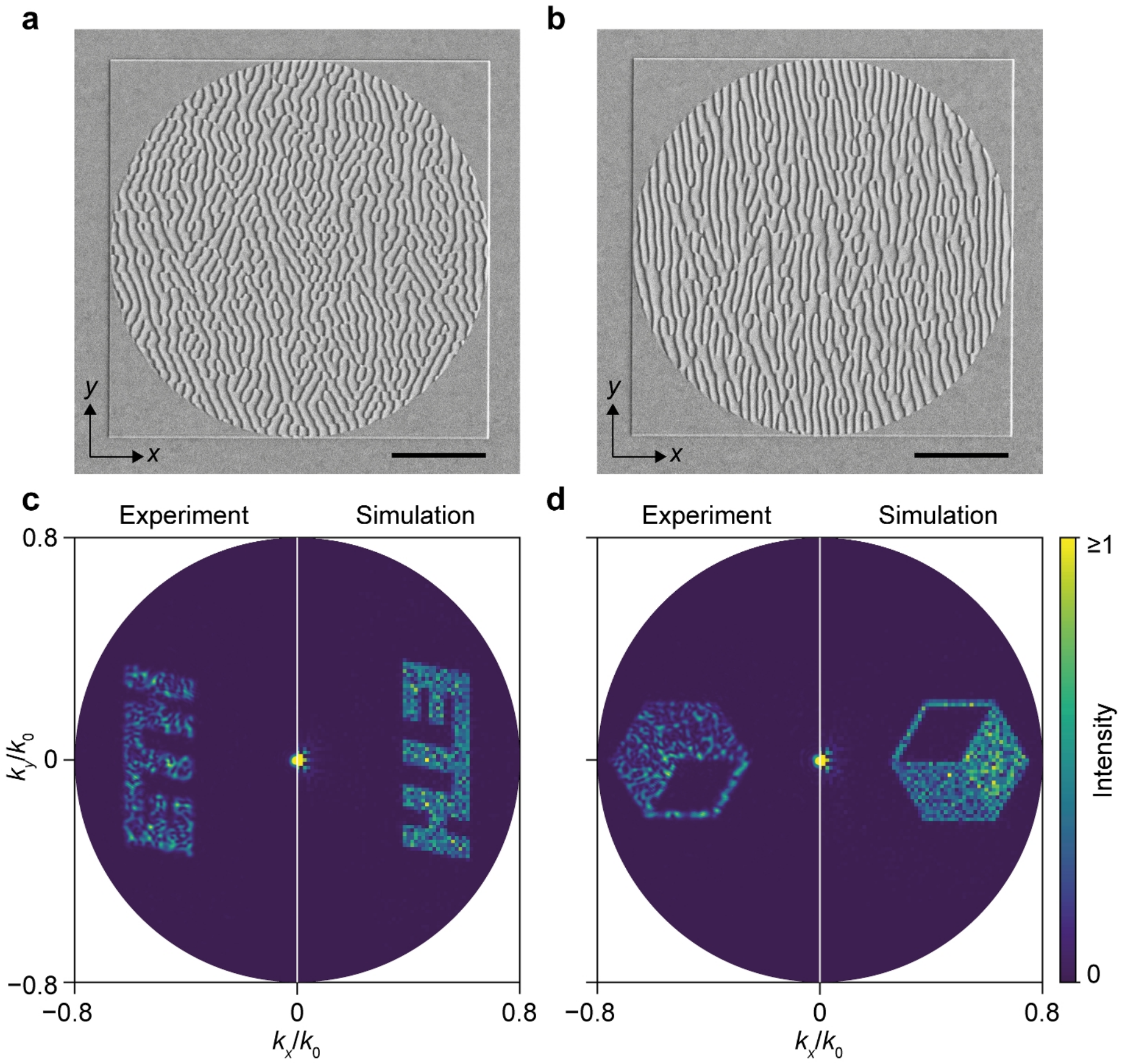

**Figure 3. Iterative design approach based on the Gerchberg–Saxton (GS) algorithm.** (a) and (b) SEM images of fabricated OFS holograms in Ag, designed with an iterative Fourier transform algorithm (IFTA) to generate the ETH logo and a grayscale cube, respectively. Both structures have a lateral size of 40 µm × 40 µm (scale bar: 10 µm) and target depths of 228 nm (ETH logo) and 236 nm (cube), tailored to incident light with $\lambda$ = 532 nm. (c) and (d) Simulated and experimentally measured far-field diffraction intensity patterns (NA of 0.8). Intensities for simulation and experiment are normalized to the respective maximum values of the diffraction image (excluding the 0$^{th}$-order specular reflection) and plotted on a linear scale. Data were acquired under the same conditions as in Figure 2 (see Supporting Information for details).

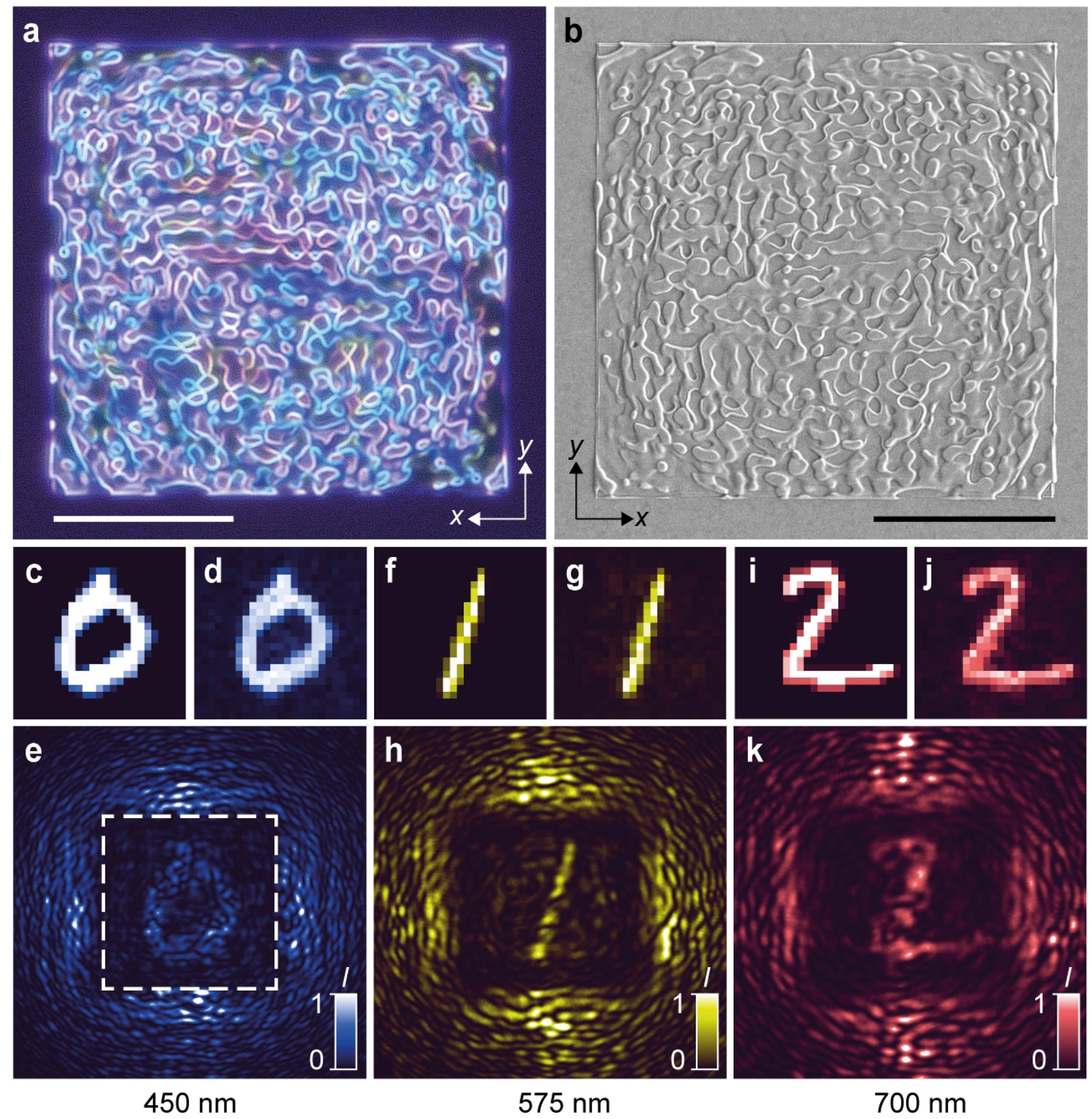

**Figure 4. Wavelength-multiplexed OFS hologram ($N_w$ = 3).** (a) Dark-field (DF) optical image of a 3-fold wavelength-multiplexed hologram in PPA, patterned with a pixel size of 40 nm × 40 nm. (b) SEM image of the corresponding structure in Ag with a lateral size of 50 µm × 50 µm and a depth of 350 nm. The structure is designed to diffract light at three different wavelengths into distinct diffraction images at 200 µm distance from the sample plane. Scale bars in (a) and (b) are 20 µm. (c),(f),(i) Target diffraction intensity patterns at the image plane for $\lambda$ = 450, 575, and 700 nm, respectively. The images correspond to handwritten digits 0, 1, and 2 from the MNIST dataset. (d),(g),(j) Simulated diffraction intensities for the respective wavelengths based on simulations using the angular-spectrum method. (e),(h),(k) Experimentally measured diffraction intensities of the diffraction images for the three different wavelengths. The dashed square in (e) indicates the field of view (FOV) of 50 µm × 50 µm for which the OFS hologram was designed. Intensities in (c)–(k) are normalized to the respective maximum values within the FOV and plotted on a linear scale. (c)–(e), (f)–(h), and (i)–(k) share the same intensity scale bars, respectively.

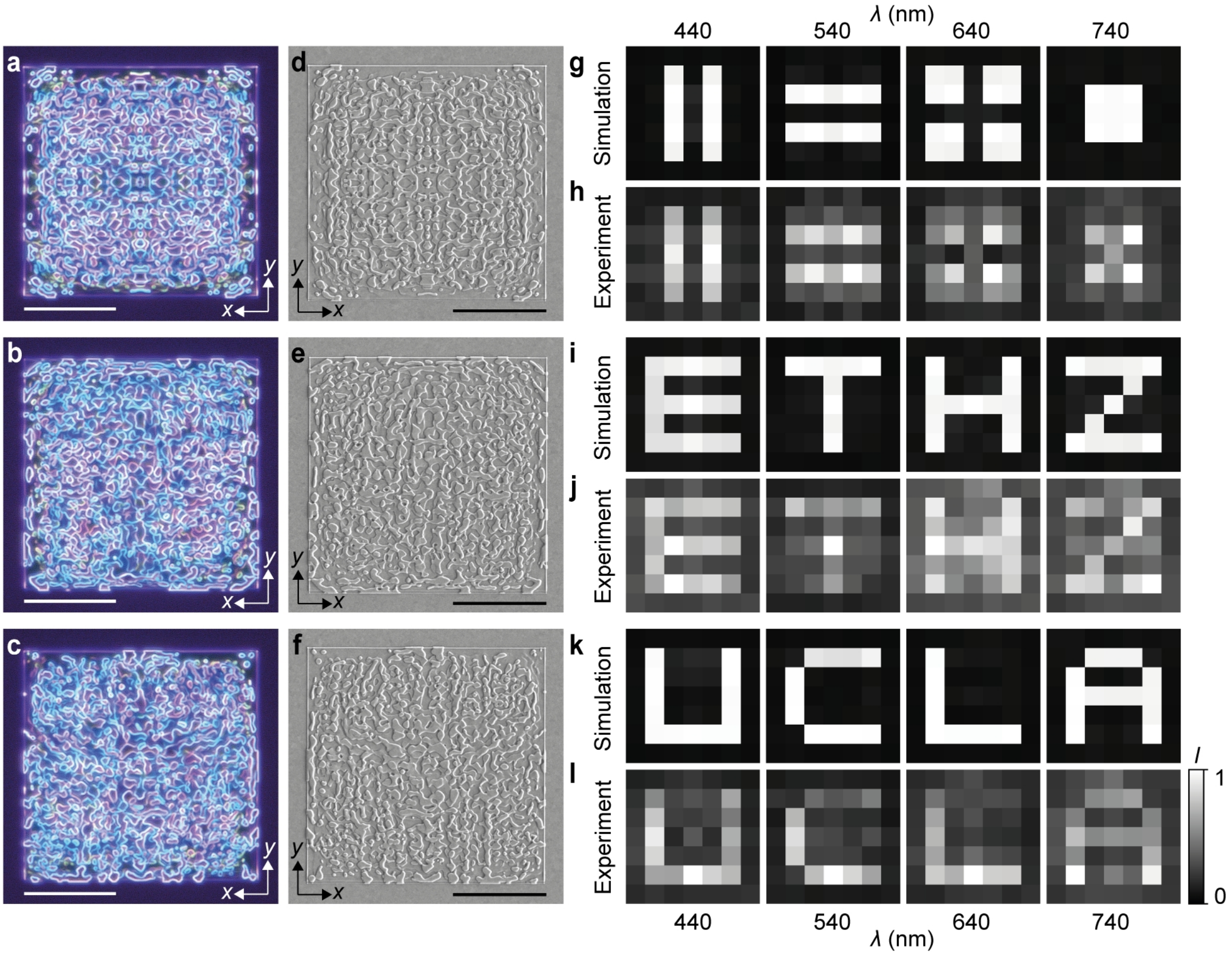

**Figure 5. Wavelength-multiplexed OFS holograms ($N_w$ = 4) with reduced image resolution.** (a)–(c) DF images of the 4-fold wavelength-multiplexed holograms in PPA for three different image series: geometric shapes (squares and rectangles), the letters "ETHZ," and the letters "UCLA", respectively. (d)–(f) SEM images of the corresponding OFS holograms fabricated in Ag. These structures are designed to diffract light at wavelengths $\lambda$ = 440, 540, 640, and 740 nm into distinct diffraction images at an image plane located 200 µm above the sample plane. They have a lateral size of 50.4 µm × 50.4 µm and a depth of 350 nm. Scale bars in (a)–(f) are 20 µm. (g)–(l) Simulated and experimentally measured diffraction intensities within the FOV, which spans 50.4 µm × 50.4 µm. The experimental data were binned to 8 × 8 pixels for direct comparison with the simulations. Each dataset in (g)–(l) was normalized to its respective maximum value and plotted on the same linear scale. The raw experimental data are provided in Figure S15 in the Supporting Information.

**Table of Contents Graphic**

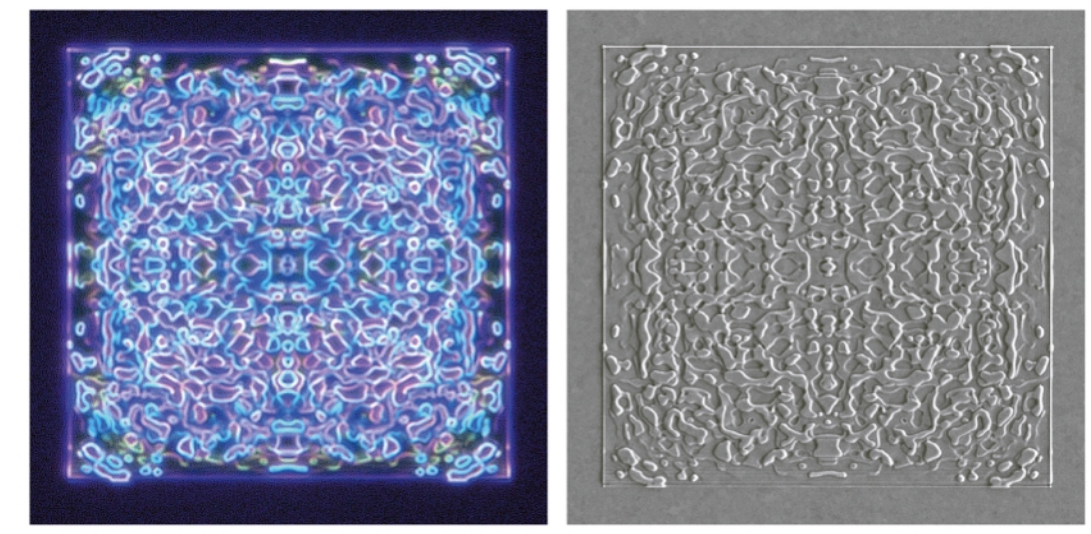

*Supporting Information for*

# Optical Fourier Surfaces for Free-Space Computer-Generated Holography

*Yannik M. Glauser,[1] Che-Yung Shen,[2,3,4] David B. Seda,[1] Çağatay Işil,[2,3,4] Patrick Benito Eberhard,[1] Hannah Niese,[1] Juri G. Crimmann,[1] Daniel Petter,[1] Gabriel Nagamine,[1] Sander J. W. Vonk,[1] Nolan Lassaline,[1,6] Aydogan Ozcan,[2,3,4,5] and David J. Norris[1,*]*

[1]Optical Materials Engineering Laboratory, Department of Mechanical and Process Engineering, ETH Zurich, 8092 Zurich, Switzerland

[2]Department of Electrical and Computer Engineering, University of California, 90095 Los Angeles, USA

[3]Department of Bioengineering, University of California, 90095 Los Angeles, USA

[4]California NanoSystems Institute (CNSI), University of California, 90095 Los Angeles, USA

[5]Department of Surgery, David Geffen School of Medicine, University of California, 90095 Los Angeles, USA

[6]Department of Physics, Technical University of Denmark, 2800 Kongens Lyngby, Denmark

**Corresponding author**

*Email: dnorris@ethz.ch

## S1. SUPPORTING DISCUSSION

### S1.1 SCALAR DIFFRACTION THEORY

The following section discusses scalar diffraction theory in the context of Fourier optics and linear system theory. It combines information from refs S1 and S2 with explanations from Goodman[S3] and Saleh and Teich.[S4]

Light can be described as an electromagnetic wave, whose propagation is governed by Maxwell's equations. In the absence of free charges and currents, these equations reduce to a set of equations for all components of the electric field **E** and the magnetic field **H**, which simplifies to a scalar-wave equation:

$$\nabla^2 u(\mathbf{r},t) - \frac{n^2}{c^2}\frac{\partial^2 u(\mathbf{r},t)}{\partial t^2} = 0\,, \tag{S1}$$

where $u(\mathbf{r},t)$ describes any field component as a function of position $\mathbf{r}$ and time $t$, $c$ is the speed of light in vacuum, and $n$ is the refractive index.

General solutions to eq S1 are monochromatic waves

$$u(\mathbf{r},t) = A(\mathbf{r})\cos[\omega t - \phi(\mathbf{r})] \tag{S2}$$

with amplitude $A(\mathbf{r})$, phase $\phi(\mathbf{r})$, and angular frequency $\omega$. Alternatively, the scalar field can be written as the real part of a complex function $U(\mathbf{r})$,

$$u(\mathbf{r},t) = \mathrm{Re}[U(\mathbf{r})\exp(-i\omega t)] \tag{S3}$$

with

$$U(\mathbf{r}) = A(\mathbf{r})\exp[i\phi(\mathbf{r})]\,. \tag{S4}$$

Inserting eq S4 into the scalar-wave equation eq S1 removes the harmonic time dependence of the monochromatic wave and leads to the Helmholtz equation

$$(\nabla^2 + k^2)U(\mathbf{r}) = 0\,. \tag{S5}$$

The wavenumber $k$ is defined by

$$k = \frac{n\omega}{c} = \frac{2\pi n}{\lambda}, \quad \text{(S6)}$$

with wavelength $\lambda$ of the light in vacuum. Among the solutions of the Helmholtz equation, the simplest set are plane waves

$$U(\mathbf{r}) = \exp(i\mathbf{k} \cdot \mathbf{r}), \quad \text{(S7)}$$

which are defined by their wavevector $\mathbf{k} = (k_x, k_y, k_z)$ with $|\mathbf{k}| = k$. For propagation along the $z$ axis, $k_z$ follows from the wavenumber $k$ and the in-plane momenta $(k_x, k_y)$ as

$$k_z = \sqrt{k^2 - {k_x}^2 - {k_y}^2}, \quad \text{(S8)}$$

which describes propagating (${k_x}^2 + {k_y}^2 \leq k^2$, real) or evanescent (${k_x}^2 + {k_y}^2 > k^2$, imaginary) waves.

Decomposing light fields into plane waves builds the foundation of Fourier optics. To obtain the spectrum of spatial frequency components $F(k_x, k_y)$ for a field $f(x, y)$, the Fourier transform is applied:

$$F(k_x, k_y) = \mathcal{F}\{f\} = \iint_{-\infty}^{+\infty} f(x, y) \exp[-i(k_x x + k_y y)] \, \mathrm{d}x\mathrm{d}y. \quad \text{(S9)}$$

Here, we use the non-unitary convention based on angular frequencies. The corresponding inverse Fourier transform is defined as

$$f(x, y) = \mathcal{F}^{-1}\{F\} = \frac{1}{4\pi^2} \iint_{-\infty}^{+\infty} F(k_x, k_y) \exp[i(k_x x + k_y y)] \, \mathrm{d}k_x\mathrm{d}k_y. \quad \text{(S10)}$$

Fourier analysis is crucial for characterizing linear systems. It is widely applied to describe light propagation through free space and linear optical components. To explain propagation along the $z$ direction from an input plane ($z = 0$) to an output plane ($z = d$), we consider the field distributions $U(\mathbf{r})$ at the two parallel planes. Assuming a linear system, the output field $U(x, y, d) = f_{\text{out}}(x, y)$ is linearly related to the input field $U(\xi, \eta, 0) = f_{\text{in}}(\xi, \eta)$. According to linear system theory, such a system is fully characterized by its impulse response $h(\xi, \eta, x, y)$. For a linear space-invariant optical system, the impulse response reduces to $h(x-\xi, y-\eta)$, depending only on relative coordinates. The output field $f_{\mathrm{out}}(x,y)$ can then be written as the convolution integral:

$$f_{\mathrm{out}}(x,y) = \iint_{-\infty}^{+\infty} f_{\mathrm{in}}(\xi,\eta) h(x-\xi, y-\eta)\, \mathrm{d}\xi \mathrm{d}\eta\,. \tag{S11}$$

By applying the convolution theorem of the Fourier transform, a linear space-invariant optical system can equivalently be described by its transfer function $H(k_x, k_y)$, which is the Fourier transform of $h(x-\xi, y-\eta)$. It simply relates the Fourier-transformed input and output signals as a multiplication:

$$F_{\mathrm{out}}(k_x, k_y) \;=\; F_{\mathrm{in}}(k_x, k_y)\, H(k_x, k_y)\,. \tag{S12}$$

For many applications, it is more convenient to perform propagation calculations in Fourier space (eq S12). In this case, the wavefront $f_{\mathrm{in}}(x,y)$ is decomposed into its angular frequency components $F_{\mathrm{in}}(k_x, k_y)$. These components are propagated by multiplication with the transfer function in Fourier space $H(k_x, k_y)$ and then transformed back to real space by the inverse Fourier transform. This approach is known as the angular spectrum method.

For propagation of light along the $z$ axis, the impulse response $h(x-\xi, y-\eta)$, based on the first Rayleigh-Sommerfeld solution, is defined as

$$h(x-\xi, y-\eta) = \frac{1}{2\pi}\frac{d}{r}\left(\frac{1}{r} - jk\right)\frac{\exp(ikr)}{r}\,, \tag{S13}$$

where

$$r = \sqrt{(x-\xi)^2 + (y-\eta)^2 + d^2}\,. \tag{S14}$$

The corresponding transfer function $H(k_x, k_y)$ is given by

$$H(k_x, k_y) = \exp(ik_z d) = \exp\left(i\sqrt{k^2 - k_x^2 - k_y^2}\, d\right), \tag{S15}$$

where we applied the definition of $k_z$ from eq S8. When the distance $d$ between the input and output plane is large compared to the wavelength $\lambda$, that is $d \gg \lambda$, the impulse response in eq S13 simplifies to

$$h(x-\xi, y-\eta) = \frac{d}{i\lambda}\frac{\exp(ikr)}{r^2}\,. \tag{S16}$$

Additional approximations can be introduced, leading to the Fresnel and Fraunhofer diffraction regimes. The Fresnel approximation is based on the Taylor expansion of eq S14. Under the assumption that

$$\max\left\{\frac{\pi}{4\lambda}[(x-\xi)^2+(y-\eta)^2]^2\right\} \ll d^3\,, \tag{S17}$$

only the first two orders of the expansion are retained, resulting in

$$r = d\left[1+\frac{1}{2}\left(\frac{x-\xi}{d}\right)^2+\frac{1}{2}\left(\frac{y-\eta}{d}\right)^2\right]. \tag{S18}$$

Substituting eq S18 into eq S16 yields

$$h(x-\xi, y-\eta) = \frac{e^{jkd}}{i\lambda d}\exp\left\{\frac{ik}{2d}[(x-\xi)^2+(y-\eta)^2]\right\}. \tag{S19}$$

The corresponding transfer function in the Fresnel regime is

$$H(k_x, k_y) = e^{ikd}\exp\left[-\frac{i\lambda d}{4\pi}\left(k_x^2+k_y^2\right)\right]. \tag{S20}$$

The full convolution integral of eq S11 can then be expressed for the Fresnel approximation by inserting eq S19:

$$f_{\mathrm{out}}(x,y) = \frac{e^{jkd}}{i\lambda d}\iint_{-\infty}^{+\infty} f_{\mathrm{in}}(\xi,\eta)\exp\left\{\frac{ik}{2d}[(x-\xi)^2+(y-\eta)^2]\right\}\mathrm{d}\xi\mathrm{d}\eta\,, \tag{S21}$$

which is known as the Fresnel diffraction integral. The Fresnel approximation restricts the propagation of light to small angles relative to the $z$ axis, meaning $k_x, k_y \ll k$. Therefore, the Fresnel and paraxial approximations are often considered equivalent, defining the Fresnel diffraction regime.

The Fraunhofer approximation introduces an even stronger condition:

$$\max\left[\frac{k(\xi^2+\eta^2)}{2}\right] \ll d\,. \tag{S22}$$

Applying eq S22 to eq S21 yields

$$f_{\mathrm{out}}(x,y) = \frac{e^{ikd}e^{i\frac{k}{2d}(x^2+y^2)}}{i\lambda d}\iint_{-\infty}^{+\infty} f_{\mathrm{in}}(\xi,\eta)\exp\left[-i\frac{k}{d}(x\xi+y\eta)\right]\mathrm{d}\xi\mathrm{d}\eta\,, \tag{S23}$$

which involves the Fourier transform of the input signal $U(\xi,\eta,0) = f_{\mathrm{in}}(\xi,\eta)$ with an additional phase term. This becomes more apparent by rewriting eq S23 as

$$f_{\text{out}}(x,y) = \frac{e^{ikd}e^{i\frac{k}{2d}(x^2+y^2)}}{i\lambda d} F_{\text{in}}\left(\frac{kx}{d}, \frac{ky}{d}\right). \tag{S24}$$

The Frauenhofer regime corresponds to the far field. In this context, the far field corresponds to Fourier space because the wave propagation can be directly described by a Fourier transform.

Similar considerations can be made for more complex linear optical systems. For example, lenses can be integrated as optical elements within such systems. As described in the Methods, we employ a $2f$-system in our optical setup to measure the far-field response of our optical Fourier surface (OFS) holograms. In this configuration, the lens performs the Fourier transform of the input signal $U(\xi,\eta,0) = f_{\text{in}}(\xi,\eta)$ without an additional spatially dependent phase term, requiring only the less restrictive Fresnel approximation.

### S1.2 PARAXIAL DIFFRACTION MODEL

Based on scalar diffraction theory, the interaction of an incoming reference wave $U_0(x,y)$ with an OFS hologram in silver (Ag) can be expressed through its complex transparency $t(x,y)$:

$$U(x,y) = U_0(x,y)t(x,y)\,. \tag{S25}$$

$U(x,y)$ defines the wavefront directly after reflection. As a phase-only hologram, the OFS locally modulates the phase of the reference wavefront by introducing an optical path difference (OPD). The transparency reads

$$t(x,y) = e^{i\Delta\phi} = e^{i\frac{2\pi}{\lambda}\text{OPD}} = e^{ik\text{OPD}} \tag{S26}$$

with additional local phase $\Delta\phi$. The OPD directly relates to the height profile $h_{\text{p}}(x,y)$ of the OFS hologram through

$$\text{OPD} = -2h_{\text{p}}(x,y)\,. \tag{S27}$$

Here, we employ a paraxial diffraction model. It assumes that light enters and exits the surface profile along the *z* axis, introducing a factor of 2 in the OPD (in and out). The sign convention is chosen such that a negative surface height corresponds to a positive phase shift (i.e., a longer propagation distance).

### S1.3 FUNDAMENTAL BUILDING BLOCKS

**Sinusoidal and blazed gratings.** We consider sinusoidal gratings as intuitive building blocks for far-field holography. Their height profile is defined by

$$h_{\mathrm{p}}(x,y) = A\sin(gx-\varphi)\,, \tag{S28}$$

with amplitude $A$, phase $\varphi$, and spatial frequency $g = \frac{2\pi}{\Lambda}$ (grating momentum), determined by the period $\Lambda$. By applying the Jacobi–Anger expansion, the sinusoidal phase term in the transparency can be approximated by

$$t(x,y) = e^{-j\,2kA\sin(gx-\varphi)} = \sum_{m=-\infty}^{\infty} J_m(-2kA)e^{im(gx-\varphi)}\,, \tag{S29}$$

where $J_m$ is the Bessel function of the first kind of order $m$. Based on eqs S24 and S29, the far-field response $f_{\mathrm{out}}(x,y)$ of an infinite sinusoidal grating is

$$f_{\mathrm{out}}(x,y) \propto \sum_{m=-\infty}^{\infty} J_m(-2kA)e^{-im\varphi}\,\delta(k_x - mg, k_y), \tag{S30}$$

where $k_x = \frac{kx}{d}$, $k_y = \frac{ky}{d}$ for distance $d$. The diffraction pattern forms a Dirac delta comb, with individual diffraction spots shifted by $g$ with respect to each other in Fourier space. The amplitude of each spot is determined by the Bessel coefficient $J_m$, which depends on the diffraction order $m$, wavenumber $k$, and amplitude $A$. The phase of each spot is determined by $-m\varphi$.

Following eq 1 of the main text, the superposition of two sinusoidal gratings with arbitrary in-plane wavevectors, $\mathbf{g}_1 = (g_{1,x}, g_{1,y})$ and $\mathbf{g}_2 = (g_{2,x}, g_{2,y})$, produces a convolution of two delta combs, yielding diffraction spots at positions $m\mathbf{g}_1 + n\mathbf{g}_2$ ($m, n \in \mathbb{Z}$):

$$\left[\sum_{m=-\infty}^{\infty} \delta(k_x - mg_{1,x}, k_y - mg_{1,y})\right] * \left[\sum_{n=-\infty}^{\infty} \delta(k_x - ng_{2,x}, k_y - ng_{2,y})\right]$$

$$= \sum_{m=-\infty}^{\infty}\sum_{n=-\infty}^{\infty} \delta[k_x - (mg_{1,x} + ng_{2,x}), k_y - (mg_{1,y} + ng_{2,y})]$$

$$= \sum_{m=-\infty}^{\infty}\sum_{n=-\infty}^{\infty} \delta[\mathbf{k} - (m\mathbf{g}_1 + n\mathbf{g}_2)] \tag{S31}$$

with $\mathbf{k} = (k_x, k_y)$.This principle is exploited for generating arbitrary diffraction images.

In contrast, a blazed grating is defined by a surface profile

$$h_\mathrm{p}(x,y) = \mathrm{mod}\left(\frac{\lambda}{2\Lambda}x, A\right), \tag{S32}$$

with surface period $\Lambda$, wavelength $\lambda$ and amplitude $A$. We apply the modulo operator, which represents division with remainder such that the surface height repeats within the range $[0, A)$. Here, the corrugation axis is along $x$, but it could be set to any in-plane orientation.

If $A$ is a multiple of half the wavelength, the transparency of the blazed grating reduces to

$$t(x,y) = e^{-i2k\cdot\mathrm{mod}\left(\frac{\lambda}{2\Lambda}x,\frac{\lambda}{2}\right)gx} = e^{-i2\frac{2\pi}{\lambda}\frac{\lambda}{2\Lambda}x} = e^{-igx}, \tag{S33}$$

where eq S32 was applied in eqs S26 and S27. The grating momentum $g = \frac{2\pi}{\Lambda}$. The far-field diffraction response $f_\mathrm{out}(x,y)$ corresponds to a single diffraction spot

$$f_\mathrm{out}(x,y) \propto \delta(k_x - g, k_y), \tag{S34}$$

with $k_x = \frac{kx}{d}, k_y = \frac{ky}{d}$. The superposition of two arbitrary blazed gratings with $\mathbf{g}_1$ and $\mathbf{g}_2$ simply results in another blazed grating with $\mathbf{g} = \mathbf{g}_1 + \mathbf{g}_2$, still diffracting light into only one spot:

$$\delta(\mathbf{k} - \mathbf{g}_1) * \delta(\mathbf{k} - \mathbf{g}_2) = \delta[\mathbf{k} - (\mathbf{g}_1 + \mathbf{g}_2)] = \delta(\mathbf{k} - \mathbf{g}) \tag{S35}$$

with $\mathbf{k} = (k_x, k_y)$. Therefore, superposing blazed gratings adds no optical functionality, making them unsuitable as building blocks for far-field holograms. It is worth noting that blazed gratings diffract light into a single spot only at the specific wavelength for which $A$ is a multiple of $\lambda/2$. This makes them spectrally selective, whereas sinusoidal gratings exhibit broadband capability.

**Sinusoidal and blazed lenses.** Analogously, we can use sinusoidal lenses as intuitive building blocks for holograms that generate desired diffraction responses at an image plane located a distance $d = f$ from the sample. In cylindrical coordinates, the surface profile is of the form

$$h_\mathrm{p}(\rho,\theta) = A\sin\left(\frac{\pi}{\lambda f}\rho^2 - \varphi\right) \tag{S36}$$

with amplitude $A$, focal distance $f$, wavelength $\lambda$ and phase $\varphi$. Applying the Jacobi–Anger expansion (as in eq S29), it can be shown that a sinusoidal lens has multiple foci at distances $\pm f/n$ with $n \in \mathbb{N}^+$, as experimentally verified in Figure S3c.

In contrast, a blazed Fresnel lens exhibits a surface profile

$$h_\mathrm{p}(\rho,\theta) = \mathrm{mod}\left(\frac{1}{4f}\rho^2, A\right). \tag{S37}$$

Similar to the blazed grating, if $A$ is a multiple of half the wavelength, the transparency of a blazed Fresnel lens reads

$$t(\rho,\theta) = e^{-i2k\cdot\mathrm{mod}\left(\frac{1}{4f}\rho^2,\frac{\lambda}{2}\right)gx} = e^{-i2\frac{2\pi}{\lambda}\cdot\frac{1}{4f}\rho^2} = e^{-i\frac{\pi}{\lambda f}\rho^2}. \tag{S38}$$

The lens introduces a parabolic phase modulation with a single focus at distance $f$, as experimentally shown in Figure S3d. The superposition of blazed Fresnel lenses with varying lateral shifts or focal distances simply results in another lens with a parabolic phase profile, providing no additional functionality. This is in strong contrast to the superposition of sinusoidal lenses, which enables richer control over the optical response, while exhibiting broadband capability. Further details are provided in Moreno *et al.*[S5]

## S1.4 SUBWAVELENGTH DIFFRACTIVE FEATURES

To investigate the influence of subwavelength surface features on the diffraction response of a phase-only hologram, we analyzed a double-sinusoidal profile of the form

$$h_\mathrm{p}(x,y) = A_1\sin(gx) + A_n\sin(ngx-\varphi), \tag{S39}$$

as illustrated in Figure S4a. The first term represents the fundamental harmonic with amplitude $A_1$ and spatial frequency $g = \frac{2\pi}{\Lambda}$, where the fundamental surface period is $\Lambda$. The second term defines the superposed higher harmonic of order $n$, with amplitude $A_n$, spatial frequency $ng$, and relative phase shift $\varphi$. Following the discussion in the Supporting Information of ref S1, the diffraction efficiency of order $l$ of a double-sinusoidal OFS, based on the paraxial model, is defined by

$$\eta_l = \left|\sum_{m=-\infty}^{\infty} J_{l-nm}(-2kA_1)J_m(-2kA_n)e^{-im\varphi}\right|^2, \tag{S40}$$

with the wavenumber of light $k = \frac{2\pi}{\lambda}$. For the discussion in Figure S4, we examine the +1st-order diffraction efficiency. We fix $l$ = +1, $\lambda$ = 550 nm, $\Lambda$ = 600 nm, $A_1$ = 50 nm, and $\varphi$ = 0, while varying $A_n$ = 0–50 nm for $n$ = 2–5. We also account for the fact that only diffraction orders inside the light cone contribute to the total diffraction efficiency. Therefore, $\eta_{+1}$ is additionally normalized by a factor *K*, representing the efficiency ratio between all diffraction orders and those that propagate to the far-field ($l$ = −1, 0, +1).[S1,S6,S7] Although the superposed higher harmonic has a period that is smaller than the wavelength of light ( $\frac{\Lambda}{n} < \lambda$ ), corresponding to non-propagating evanescent waves, it still contributes to all the other orders via the convolution principle discussed above (see eq S40 and Figure S4).

## S1.5 LINEAR DESIGN APPROACH

The linear approach is based on the first-order Taylor expansion

$$e^{i\phi} = 1 + i\phi + \mathcal{O}[(i\phi)^2] \,, \tag{S43}$$

where we neglect all higher-order terms of phase $\phi$. Here $\phi$ represents the phase modulation as a function of position, $\phi(x, y)$. Applying the Fourier transform to calculate the far-field response leads to

$$\mathcal{F}\{e^{i\phi}\} \approx \mathcal{F}\{1 + i\phi\} \; = \; 4\pi^2\delta(0) + i\mathcal{F}\{\phi\} \tag{S44}$$

with $\delta(0)$ representing the specular reflection ($0^{\text{th}}$-order diffraction) in the center of Fourier space. The only tunable part of the diffraction response is represented by $\mathcal{F}\{\phi\}$. Because the phase shift $\phi$, which is introduced by the surface profile of our OFS hologram, is real-valued, $\mathcal{F}\{\phi\}$ must be Hermitian:

$$\mathcal{F}\{\phi\}(-k_x, -k_y) = \mathcal{F}\{\phi\}^*(k_x, k_y) \,. \tag{S45}$$

We take this symmetry constraint into account during the linear design process, when selecting the target diffraction image.

This linear approach has also been applied for diffractive structures that interact with guided plasmonic and photonic modes. In that case, the additional in-plane momentum of the incident wave was anticipated in the design of the surface profile.[S2]

## S1.6 (QUASI)CRYSTALS

Azimuthally symmetric superpositions of sinusoidal profiles lead to crystalline ($n$ = 1, 2, 3) or quasicrystalline structures ($n \geq 4$) of the form

$$h_{\mathrm{p}}(x,y) = \frac{A}{n}\sum_{i=0}^{n-1} \cos\left\{g\left[x\cos\left(\frac{\pi}{n}i\right) + y\sin\left(\frac{\pi}{n}i\right)\right]\right\}, \tag{S41}$$

with fixed amplitude $A$ and spatial frequency $g$. In the limit of infinitely many sinusoidal components ($n \to \infty$), the surface profile becomes rotationally invariant. In cylindrical coordinates, it can be written as

$$h_{\mathrm{p}}(\rho,\theta) = A\,J_0(g\rho) = \frac{A}{\pi}\int_0^{\pi} \cos[g\rho\sin(\tau)]\,d\tau\,, \tag{S42}$$

where $J_0$ defines the Bessel function of the first kind and zeroth order. The far-field diffraction response of this profile produces a ring of radius $g$ in Fourier space (see Figure S6j).

## S1.7 ITERATIVE FOURIER TRANSFORM ALGORITHM

In this work, we applied a generalized Gerchberg–Saxton (GS) algorithm, schematically illustrated in Figure S11. The method iteratively transitions between the object domain (sample plane) and Fourier domain (image plane, far field), imposing constraints on the wavefronts in both domains. In the following discussion, we describe our iterative approach in the context of the original GS algorithm.[S8] Where more general capabilities are introduced through recent modifications, we refer to the relevant literature.

The iteration procedure starts with an initial guess $\varphi_0$ for the phase, which we randomly initialize between 0 and $2\pi$. We consider a reference wave $r(x,y)$ with a uniform amplitude $|r|$ and phase $\varphi_{\mathrm{ref}}$ = 0. The amplitude profile is a circular spot, accounting for the fact that we collect light from the OFS holograms using a circular aperture in the collection path of our setup (see Figure S2a). Combining the reference wave with the phase profile $\varphi_0$ leads to the initial wavefront in the object domain $g_0$:

$$g_0 = |f|\,e^{i\varphi_0}\,. \tag{S46}$$

During each iteration $n$, the algorithm goes through 4 steps (see Figure S11):

- **Step 1.** Calculate the corresponding field in the Fourier domain using the Fourier transform:

$$\mathcal{F}\{g_n\} \longrightarrow G_n(k_x, k_y) = |G_n| e^{i\Phi_n} \,. \tag{S47}$$

The Fourier transform of $g_n$ yields $G_n$ with amplitude and phase, $|G_n|$ and $\Phi_n$, respectively. $G_n$ represents the unconstrained far-field response in iteration $n$.

- **Step 2.** Apply Fourier domain constraints:

$$|F|\, e^{i\Phi_n} \longrightarrow G_n'(k_x, k_y) = |G_n'| e^{i\Phi_n} \,. \tag{S48}$$

The amplitude of the far-field response is set to the target amplitude image $|F|$, while the phase remains unchanged.

We adapted this step by dividing the Fourier domain into a constrained signal region and an unconstrained noise region.[S9] By allowing amplitude noise in irrelevant areas of the Fourier domain, we provide more design freedom for the OFS hologram to generate the desired diffraction image in the signal region. This approach is referred to as mixed-region amplitude freedom (MRAF). It introduces a weight parameter $m$ between 0 and 1, which we set to $m = 0.3$ for our designs (see Pasienski and DeMarco[S9] for a detailed discussion).

- **Step 3.** Take the inverse Fourier transform back to the object domain:

$$\mathcal{F}^{-1}\{G_n'\} \longrightarrow g_n'(x, y) = |g_n'| e^{i\varphi_n'} \,. \tag{S49}$$

The inverse Fourier transform of $G_n'$ yields $g_n'$, with amplitude and phase, $|g_n'|$ and $\varphi_n'$, respectively. $g_n'$ represents the unconstrained wavefront in the object domain after iteration $n$.

- **Step 4.** Apply amplitude constraints in the object plane:

$$|r| e^{i\varphi_n'} \longrightarrow g_{n+1}(x, y) = |g_{n+1}| e^{i\varphi_{n+1}} \,. \tag{S50}$$

The amplitude of the reference wave is set to the circular reference wave $|r|$, while the phase remains unchanged.

We adapted this step by employing the input–output approach by Fienup.[S10,S11] In this context, steps 1 to 3 are considered as a nonlinear function with input $g_n$ and output $g_n'$. In each iteration, we look for an input that produces an output that best satisfies the object-domain constraints. Similar to step 2, this approach uses a weight parameter $\beta$ between 0 and 1, which we set to $\beta$ = 0.9 for our designs (see Fienup[S10,S11] for a detailed discussion).

After several iterations ($n \leq N$ with $N$ = 100), the algorithm converges to a final phase map ($\varphi_n \rightarrow \varphi_{\text{final}}$), which is then converted to a surface profile using the paraxial diffraction model (eqs S26 and S27). We address the fabrication constraints by applying a Gaussian filter (standard deviation $\sigma$ = 60 nm) to $\varphi_{\text{final}}$ to smooth the final profile.

## S1.8 MACHINE-LEARNING-BASED INVERSE-DESIGN APPROACH FOR WAVELENGTH MULTIPLEXED HOLOGRAMS

To optimize our wavelength-multiplexed OFS designs, we employed a machine-learning-based inverse-design framework with a custom-formulated multi-objective loss function. The diffractive reflective designs were trained using stochastic gradient descent with error backpropagation to minimize a loss function $\mathcal{L}$. The loss function is composed of three sub-terms accounting for image reconstruction accuracy ($\mathcal{L}_{\text{image}}$), fabrication feasibility ($\mathcal{L}_h$), and diffraction efficiency ($\mathcal{L}_{\text{eff}}$).

The primary objective is to ensure high-fidelity image reconstruction across all wavelength channels. This is achieved using a normalized mean-squared error (MSE) loss to penalize the structural difference between the diffractive output intensity $O_w(x,y)$ and the ground truth $O_w^{\text{GT}}(x,y)$ within the field of view (FOV) at each wavelength channel $\lambda_w$. The loss $\mathcal{L}_w$ for a specific wavelength is expressed as

$$\mathcal{L}_w = \sum_{x=1}^{N_x} \sum_{y=1}^{N_y} \left| O_w^{\text{GT}}(x,y) - \sigma_w O_w(x,y) \right|^2 . \tag{S51}$$

In this equation, $\sigma_w$ is a normalization factor for the diffractive output, defined as

$$\sigma_w = \frac{\sum_{x=1}^{N_x} \sum_{y=1}^{N_y} O_w^{\text{GT}}(x,y)}{\sum_{x=1}^{N_x} \sum_{y=1}^{N_y} O_w(x,y)} , \tag{S52}$$

where $N_x$ and $N_y$ represent the number of discrete steps along the $x$ and $y$ directions within the output FOV, respectively.

During training, the total loss is computed by averaging the individual losses $\mathcal{L}_w$ over all the $N_w$ wavelength channels, ensuring simultaneous optimization for all channels. The overall loss function is then given by

$$\mathcal{L}_{\text{image}} = \sum_{w=1}^{N_w} \alpha_w \mathcal{L}_w \,, \tag{S53}$$

where $\alpha_w$ represents the dynamic channel balance weight for each $\mathcal{L}_w$, which adjusts the performance balance across different wavelength channels during the optimization.[S12] Initially set to 1 for all wavelengths $w$, $\alpha_w$ is adaptively updated during each training iteration based on the deviation of each channel loss from the mean loss $\mathcal{L}_{\text{mean}} = \sum_{w=1}^{N_w} \mathcal{L}_w$ calculated across all wavelength channels, i.e.,

$$\alpha_w \leftarrow \max[0.1(\mathcal{L}_w - \mathcal{L}_{\text{mean}}) + \alpha_w,\ 0] \,. \tag{S54}$$

To account for fabrication constraints of the diffractive surface, we added two fabrication-aware terms: thickness variation loss and Gaussian smoothing. First, we introduced a thickness variation loss $\mathcal{L}_h$ that suppresses abrupt changes between adjacent pixels in the resulting surface profile. This helps reduce potential simulation–fabrication mismatch by penalizing variations exceeding a threshold $h_{\text{thresh}}$:

$$\begin{aligned} \mathcal{L}_h = & \sum_{x=1}^{N_x^h - 1} \sum_{y=1}^{N_y^h} \max\left[\left|h_{\text{p}}(x+1, y) - h_{\text{p}}(x, y)\right| - h_{\text{thresh}}, 0\right] \\ & + \sum_{x=1}^{N_x^h} \sum_{y=1}^{N_y^h - 1} \max\left[\left|h_{\text{p}}(x, y+1) - h_{\text{p}}(x, y)\right| - h_{\text{thresh}}, 0\right] , \end{aligned} \tag{S55}$$

where $N_x^h$ and $N_y^h$ represent the number of discrete steps along the $x$ and $y$ directions for the diffractive surface, respectively.

In addition, we apply Gaussian smoothing to the diffractive surface to further enhance its spatial smoothness and reduce high-frequency variations that could hinder fabrication. This was implemented by convolving the surface height map $h_{\text{p}}(x, y)$ with a Gaussian kernel $G(x, y)$:

$$h_{\text{smoothed}}(x, y) = \left(h_{\text{p}} * G\right)(x, y) \,. \tag{S56}$$

$G(x, y)$ is the 2D Gaussian kernel, defined as

$$G(x, y) = \frac{1}{2\pi\sigma^2} \exp\left(-\frac{x^2 + y^2}{2\sigma^2}\right), \tag{S57}$$

where $\sigma$ = 320 nm was used in our designs.

To enhance the energy efficiency of our system, we also introduced a modified loss function by adding a diffraction efficiency penalty term $\mathcal{L}_{\text{eff}}$:

$$\mathcal{L}_{\text{eff}} = \eta_{\text{thresh}} - \min(\eta, \eta_{\text{thresh}}), \tag{S58}$$

$$\eta = \frac{\sum_{w=1}^{N_w} \sum_{x=1}^{N_x} \sum_{y=1}^{N_y} O_w(x, y)}{\sum_{w=1}^{N_w} \sum_{x=1}^{N_x} \sum_{y=1}^{N_y} I_w(x, y)}. \tag{S59}$$

$I_w(x, y)$ is the input intensity at the reflective diffractive surface, and $O_w(x, y)$ is the intensity at the output plane within the FOV at each wavelength channel $\lambda_w$. $\eta_{\text{thresh}}$ is the energy efficiency threshold set to maintain a desired output diffraction efficiency. To be specific, when the actual efficiency $\eta$ falls below $\eta_{\text{thresh}}$, the efficiency loss becomes non-zero, providing the necessary gradients for the model during training.

Combining all the loss components, the total loss function used for training is

$$\mathcal{L} = \mathcal{L}_{\text{image}} + \beta_h \mathcal{L}_h + \beta_{\text{eff}} \mathcal{L}_{\text{eff}}. \tag{S60}$$

For all the wavelength-multiplexed reflective surface designs reported in the main text, the values of $\beta_{\text{eff}}$ and $\eta_{\text{thresh}}$ were empirically set as 10 and 0.05, respectively. The values of $\beta_h$ and $h_{\text{thresh}}$ were empirically set as 100 and 40 nm, respectively.

We utilized the peak signal-to-noise ratio (PSNR) metric to evaluate the structural fidelity of the intensity image $O_w(x, y)$ produced by the diffractive structure against the ground truth $O_w^{\text{GT}}(x, y)$. For a specific wavelength channel $\lambda_w$, the PSNR value is computed using

$$\text{PSNR}_w = -10\log_{10}\left[\sum_x \sum_y \left|O_w^{\text{GT}}(x, y) - \sigma_w O_w(x, y)\right|^2\right]. \tag{S61}$$

## S2. SUPPORTING FIGURES

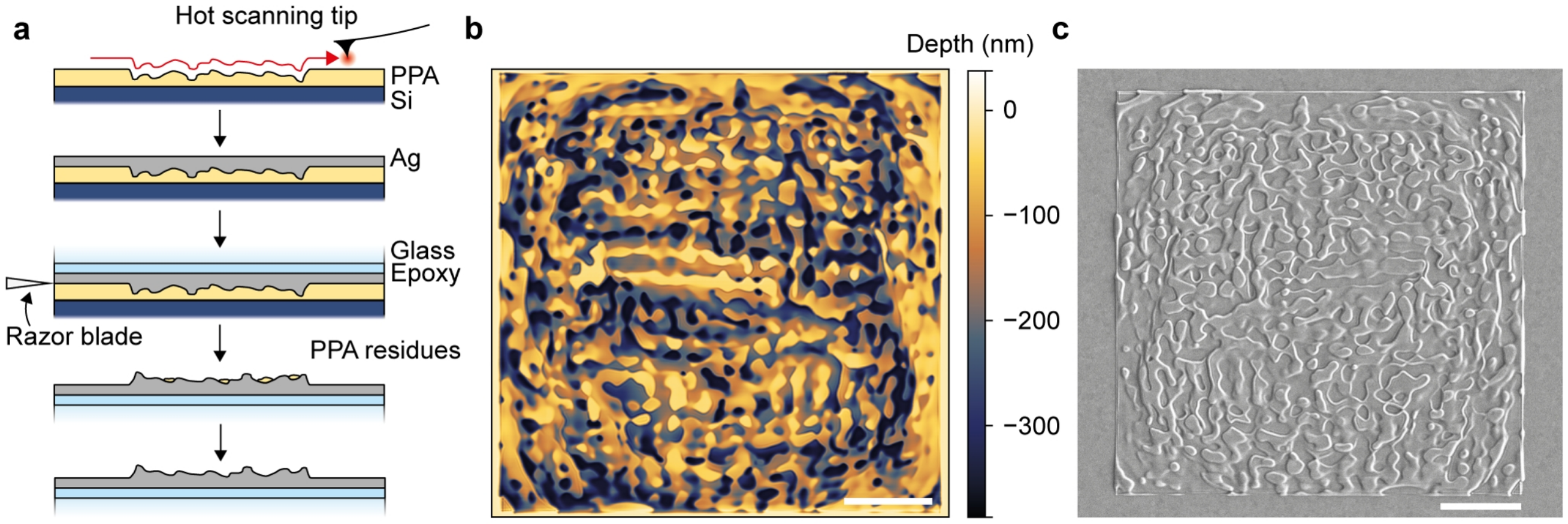

**Figure S1. Fabrication procedure of optical Fourier surfaces (OFSs).** (a) Process flow for the fabrication of an OFS, incorporating (i) thermal scanning-probe lithography (TSPL), (ii) thermal evaporation of silver (Ag), (iii) template stripping, and (iv) anisole cleaning. (b) Topography data of an OFS hologram (3-fold wavelength-multiplexed, used in Figure 4 in the main text) in poly(phthalaldehyde) (PPA) measured by the TSPL tool. (c) Scanning electron microscopy (SEM) image of the hologram transferred to Ag. The final OFSs in Ag is inverted and horizontally mirrored with respect to the structure in PPA due to template stripping. The scale bar in (b) and (c) is 10 μm.

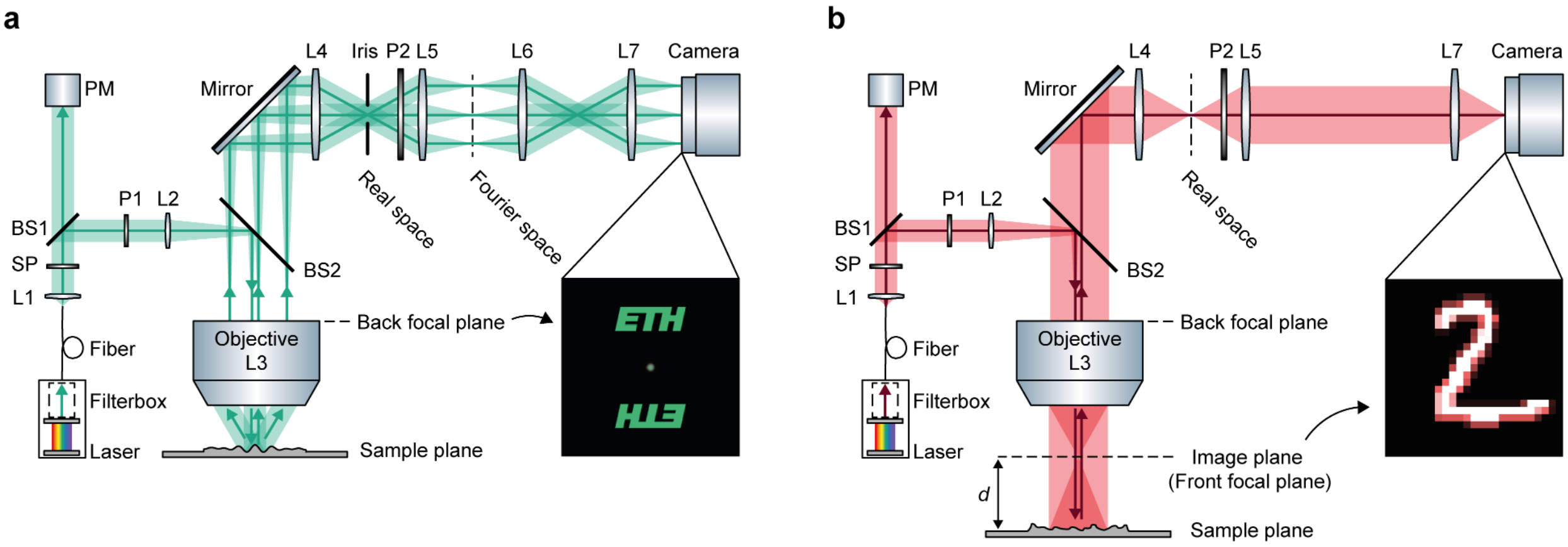

**Figure S2. Optical setup with two configurations.** (a) Far-field configuration used to image the diffraction response in the back focal plane (Fourier plane) of the objective. For these experiments, the wavelength was fixed at 532 nm. (b) Real-space configuration used to detect the diffraction response in an arbitrary image plane located at a defined distance from the sample plane. For the experiments on spectral multiplexing, the wavelength was swept accordingly. To switch from configuration (a) to (b), lens L6 and the iris were removed, and the sample plane was axially defocused by the desired distance $d$. The individual optical components of the setup are described in more detail in Methods.

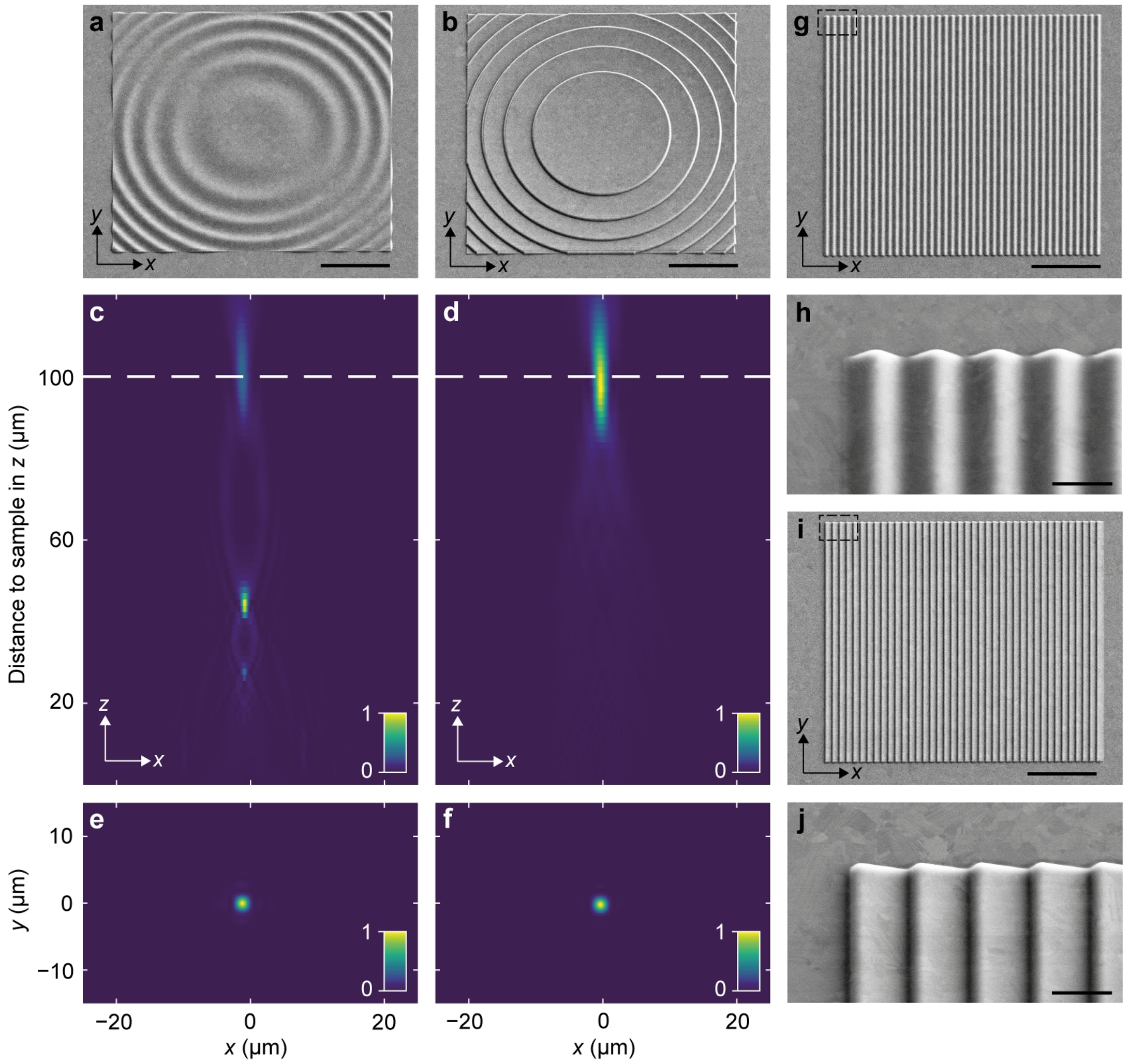

**Figure S3. Fundamental structures for holography in the far field and arbitrary image plane.** (a),(b) SEM images of a sinusoidal lens and a blazed Fresnel lens in Ag, both with focal length $f$ = 100 µm, depth of 266 nm, and design wavelength $\lambda$ = 532 nm. (c),(d) Measured intensity profiles as a function of defocus $z$, showing multiple focal spots for the sinusoidal lens and a single spot at $f$ for the blazed lens, consistent with scalar diffraction theory (see Section S1.3). (e),(f) Cross-sections of (c) and (d) at $z$ = 100 µm (white dashed lines), normalized to their respective maxima. (g),(i) SEM images of sinusoidal and blazed gratings in Ag with depths of 200 and 218 nm and period $\Lambda$ = 1000 nm. (h),(j) Zoom-ins of the dashed rectangles in (g) and (i). All SEM images were captured at a tilt angle of 30°. Scale bars are 10 µm.

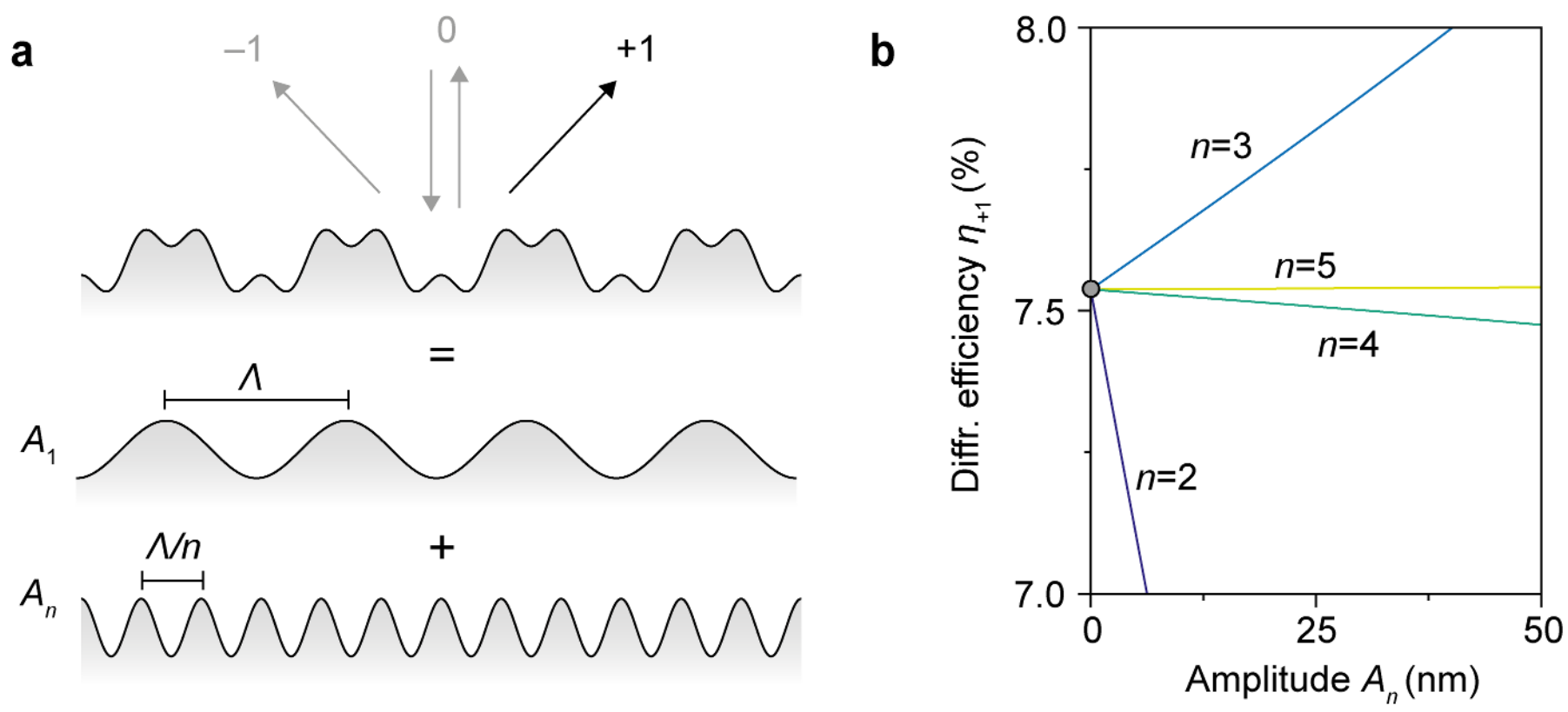

**Figure S4. Influence of high spatial frequencies with subwavelength features on the diffraction response of OFSs.** (a) Schematic of an OFS composed of a fundamental harmonic with period $\Lambda$ = 600 nm superposed with a higher harmonic of period $\Lambda/n$ ($n \geq 2$; $n$ = 3 shown here). The amplitude of the fundamental harmonic is fixed at $A_1$ = 50 nm, while the higher harmonic has a variable amplitude $A_n$ = 0–50 nm (with $A_3$ = $A_1$ shown here). The influence of the higher-order component on the +1$^{st}$ diffraction order of the OFS is evaluated. The diffraction efficiency is calculated using scalar diffraction theory (see Section S1.4), assuming a wavelength of $\lambda$ = 550 nm under normal incidence to ensure that only the 0$^{th}$, +1$^{st}$, and −1$^{st}$ diffraction orders are non-evanescent. (b) +1$^{st}$-order diffraction efficiency $\eta_{+1}$ as a function of higher-harmonic amplitude $A_n$ for $n$ = 2–5. $\eta_{+1}$ for the fundamental component alone is indicated by the gray circle. The second harmonic ($n$ = 2, purple) significantly modifies the optical response, even though this period does not directly diffract light into the 1$^{st}$ orders. Higher harmonics with $n$ = 3 (blue) and $n$ = 4 (green) induce only minor changes, while harmonics with $n$ = 5 (yellow) and above have a negligible effect, regardless of their amplitude.

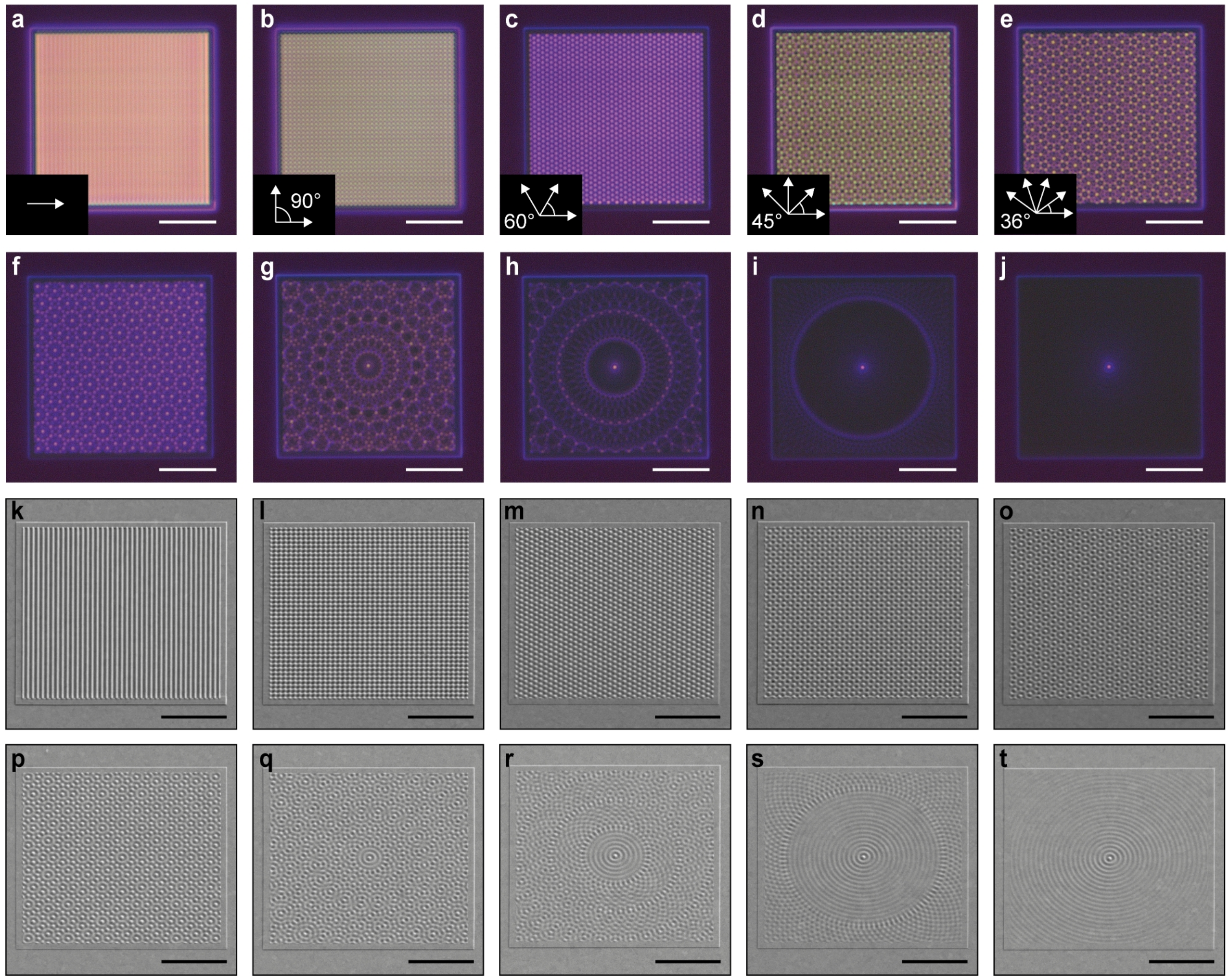

**Figure S5. (Quasi)crystalline profiles in PPA and Ag.** (a)–(j) Dark-field (DF) images of (quasi)crystalline structures in PPA composed of 1, 2, 3, 4, 5, 6, 10, 20, 50, and infinitely many symmetrically superposed sinusoids with a period of $\Lambda$ = 750 nm, respectively. The total depth is fixed at 200 nm. In the limit of infinitely many superposed sinusoids, the surface structure becomes rotationally symmetric, with a height profile described by a Bessel function $J_0$. (k)–(t) SEM images of the corresponding structures transferred to Ag. All SEM images were captured at a tilt angle of 30°. Scale bars in all subpanels are 10 μm.

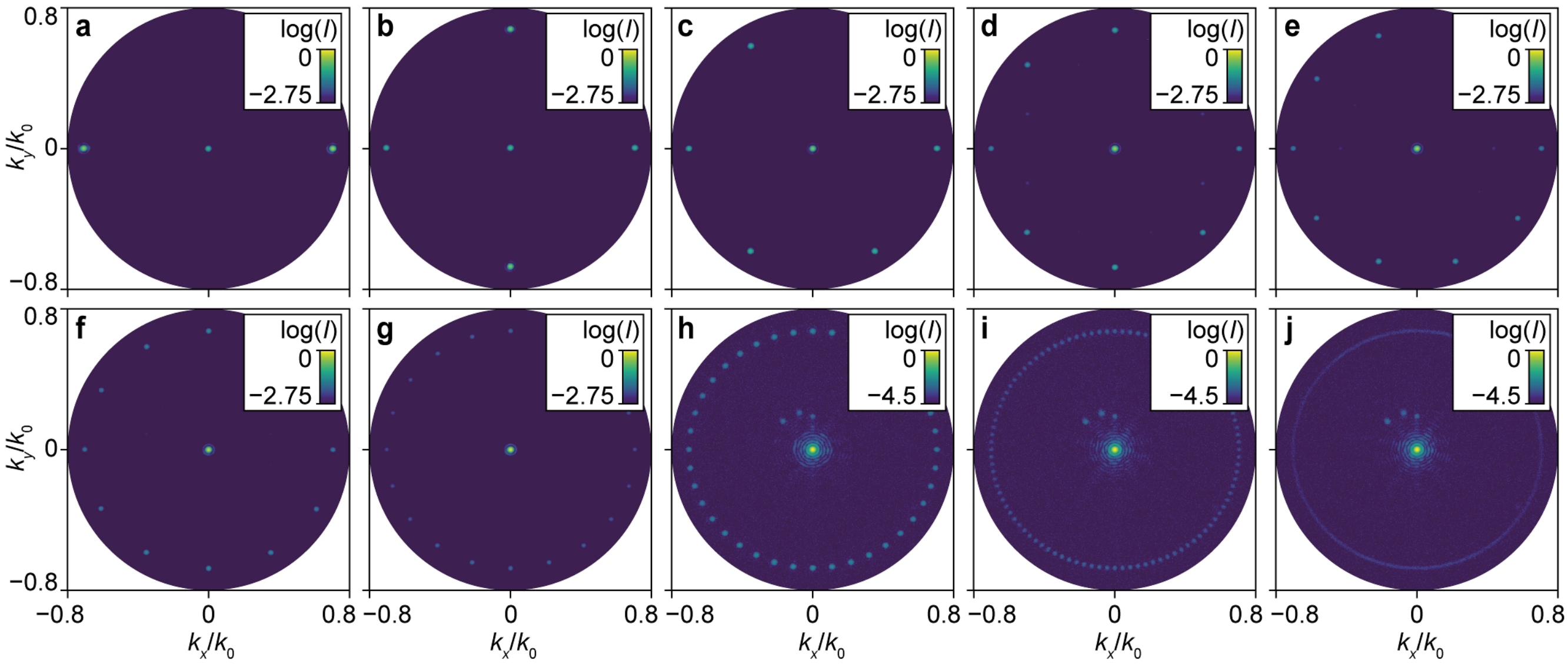

**Figure S6. Diffraction patterns of (quasi)crystalline OFSs.** (a)–(j) Measured intensity of diffraction in Fourier space (back focal plane of the objective with NA = 0.8) for the (quasi)crystalline structures shown in Figure S5k–t. Coherent light with a wavelength of $\lambda$ = 532 nm, in the form of a plane wave (i.e., a broad Gaussian beam), was used under normal incidence. The intensities are background-corrected and normalized with respect to the maximum diffraction signal of a flat Ag surface. Data are plotted on a logarithmic intensity scale. Since the intensity of the diffraction spots decreases with increasing numbers of superposed sinusoids, the color scale is adjusted in panels (f)–(j) to enhance visibility. Artifacts around the central $0^{th}$-order diffraction spot arise from back reflections within the optical setup.

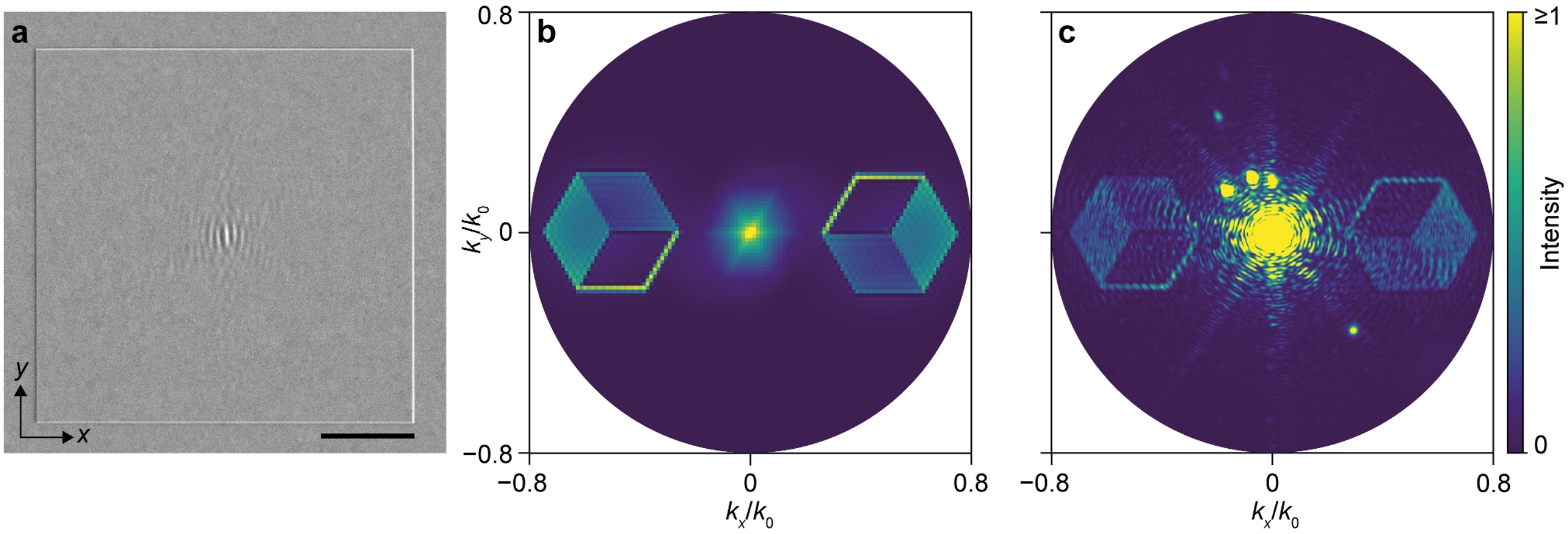

**Figure S7. Grayscale diffraction image based on the linear design approach.** (a) SEM image of an OFS hologram in Ag, designed to diffract light in the form of a cube with facets of varying intensities. The surface structure has a target depth of 266 nm. The scale bar is 10 µm. (b) Simulated and (c) experimentally measured far-field diffraction intensities in Fourier space (i.e., the back-focal plane of the objective with a numerical aperture NA of 0.8). Intensities in (b) and (c) are normalized to the respective maximum values of the diffraction image (excluding the $0^{th}$-order specular reflection) and plotted on a linear scale. The simulation is based on scalar diffraction theory. The experiment was performed under normal incidence using light with $\lambda = 532$ nm. Panels (b) and (c) correspond to the grayscale counterpart of Figure 2c,d in the main text.

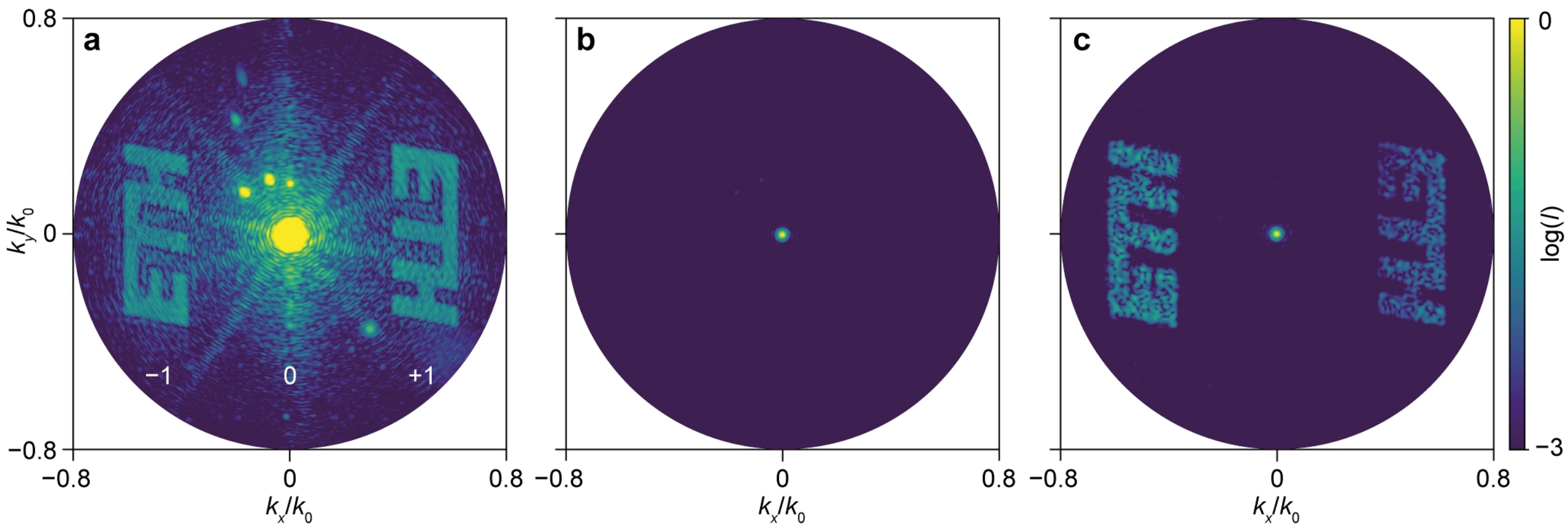

**Figure S8. Efficiency characterization for the linear design and iterative Fourier transform algorithm (IFTA).** (a) Diffraction image in the back-focal plane from an OFS hologram using the linear approach (intensity data from Figure 2d in the main text, plotted on a log scale). The ETH logo appears symmetrically as the $+1^{st}$ and $-1^{st}$ diffraction orders around the specular reflection ($0^{th}$ order). We saturate the image to clearly observe the diffraction image above the background. (b) Reference measurement on a flat Ag surface showing only the specular reflection. To extract the diffraction efficiency of the OFS hologram in (a), we integrated over the diffraction orders and used the total intensity, corrected for the different acquisition settings, from the reference measurement. We find diffraction efficiencies $\eta_{-1}^{\mathrm{L}} = 0.11\%$ and $\eta_{+1}^{\mathrm{L}} = 0.10\%$. (c) Same as in (a) but for a hologram designed through the generalized Gerchberg–Saxton (GS) algorithm. Here, we simply integrate over the diffraction orders to find the diffraction efficiencies $\eta_{-1}^{\mathrm{GS}} = 32.9\%$ and $\eta_{+1}^{\mathrm{GS}} = 15.2\%$.

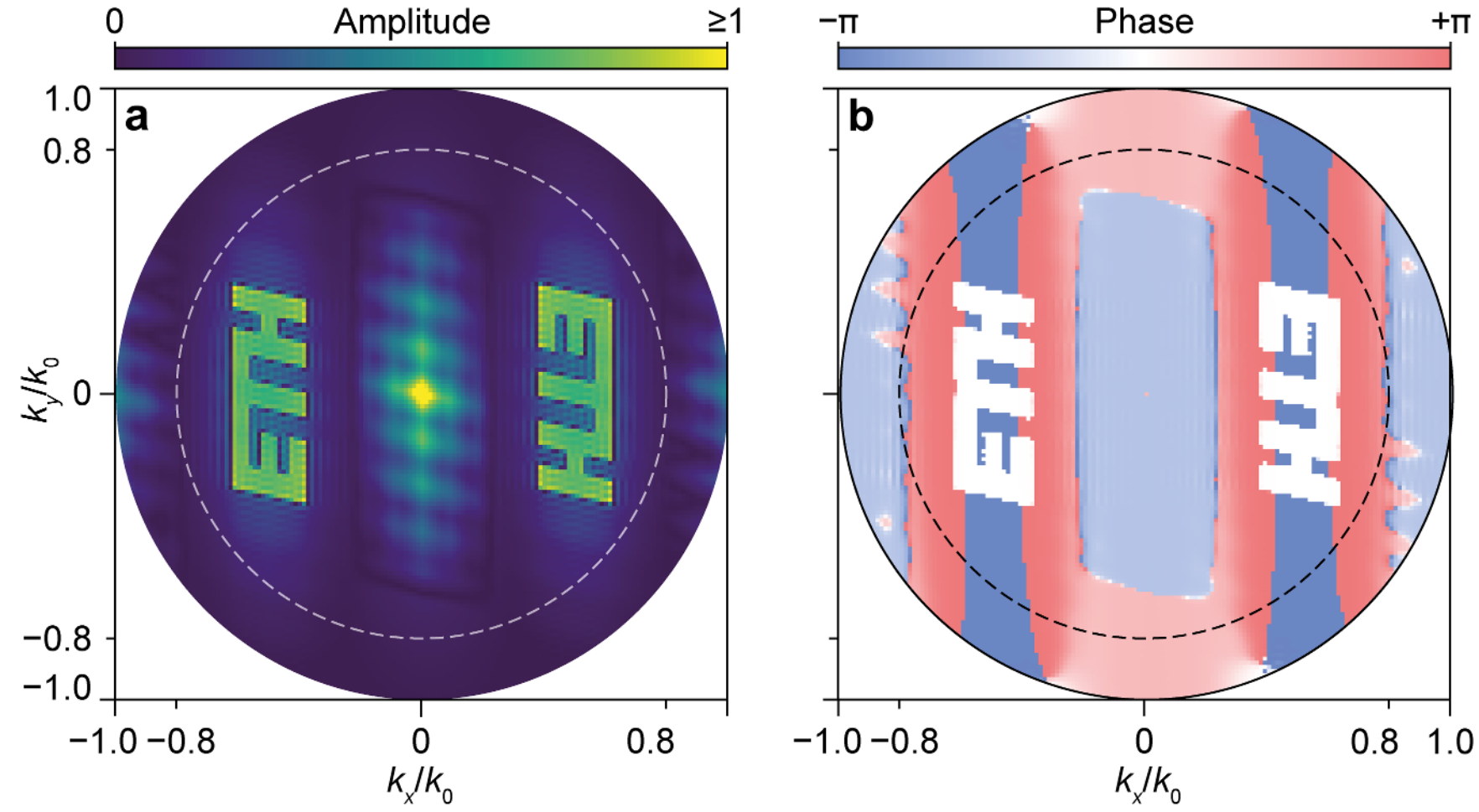

**Figure S9. Amplitude and phase control with the linear design approach.** (a) Simulated amplitude and (b) simulated phase of the complex wavefront in Fourier space for the OFS hologram depicted in Figure 2a. The numerical aperture NA = 0.8 is indicated by the dashed circle. Simulations are based on scalar diffraction theory, assuming light under normal incidence with $\lambda = 532$ nm. The amplitude in (a) is normalized to the maximum value of the diffraction image (excluding the $0^{th}$-order specular reflection) and plotted on a linear scale. The phase in (b) is uniform across the diffraction image, as intended by the design algorithm. Provided that the OFS design constraints are satisfied, in principle, any arbitrary phase map can be realized in the far-field diffraction pattern by appropriately defining the initial complex-valued target image (as shown in Figure 2a). Interestingly, a relative phase shift of approximately $\pi$ is observed between the diffraction image and background, although the phase of the background in the target image is undefined due to its zero amplitude. This phase shift introduces a node in amplitude between the diffraction image and background, facilitating sharp contours.

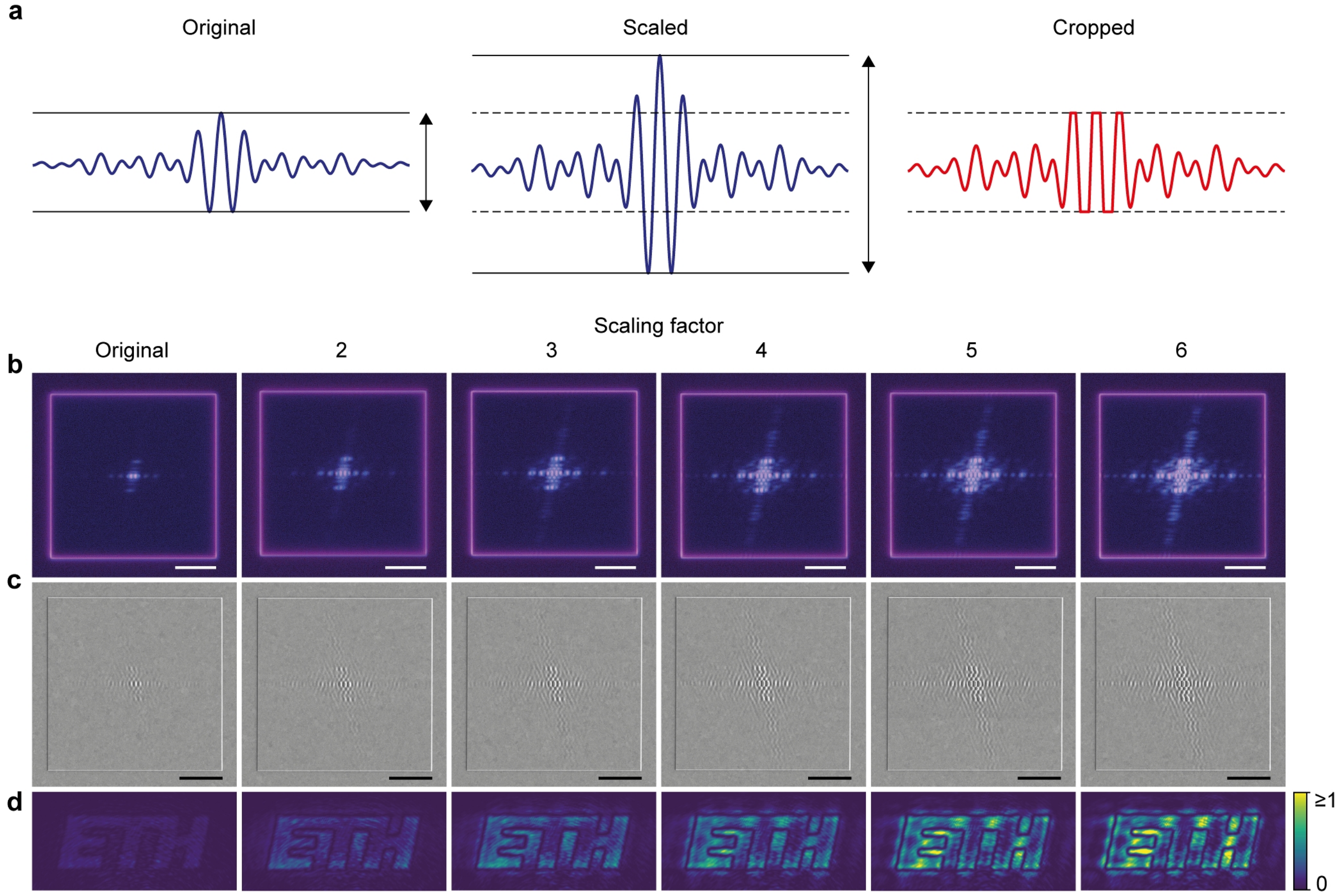

**Figure S10. Linear design approach with additional scaling and cropping of the height profile.** (a) Strategy to increase the diffracted intensity of the holograms based on the linear design approach. Their original profile (see Figure 2 in the main text) is scaled to increase the contribution of each sinusoidal component. Due to fabrication constraints, the depth of an OFS is limited. We therefore investigate how the diffraction signal is affected when the overall profile is linearly expanded in the $z$ direction and cropped to maintain a fixed depth of 266 nm. In the linear design approach, the full depth range is typically used only in the center of the structure, which occupies a small portion of the total area, while most other regions remain shallow. (b) DF images of the scaled and cropped OFS holograms in PPA. The scaling factors range from 1–6, respectively. With increasing scaling factor, the structure becomes more binary in the center, which leads to more scattered light. (c) SEM images of the corresponding OFS holograms in Ag, fabricated from the structures in (b). (d) Diffracted far-field intensity pattern for the structures in (c). Intensities in (d) are normalized to the maximum value of the diffraction image with a scaling factor of 4 (excluding the $0^{th}$-order specular reflection) and plotted on a linear scale. The experiment was performed under normal incidence using light with $\lambda = 532$ nm. Scale bars in (b) and (c) are 10 µm.

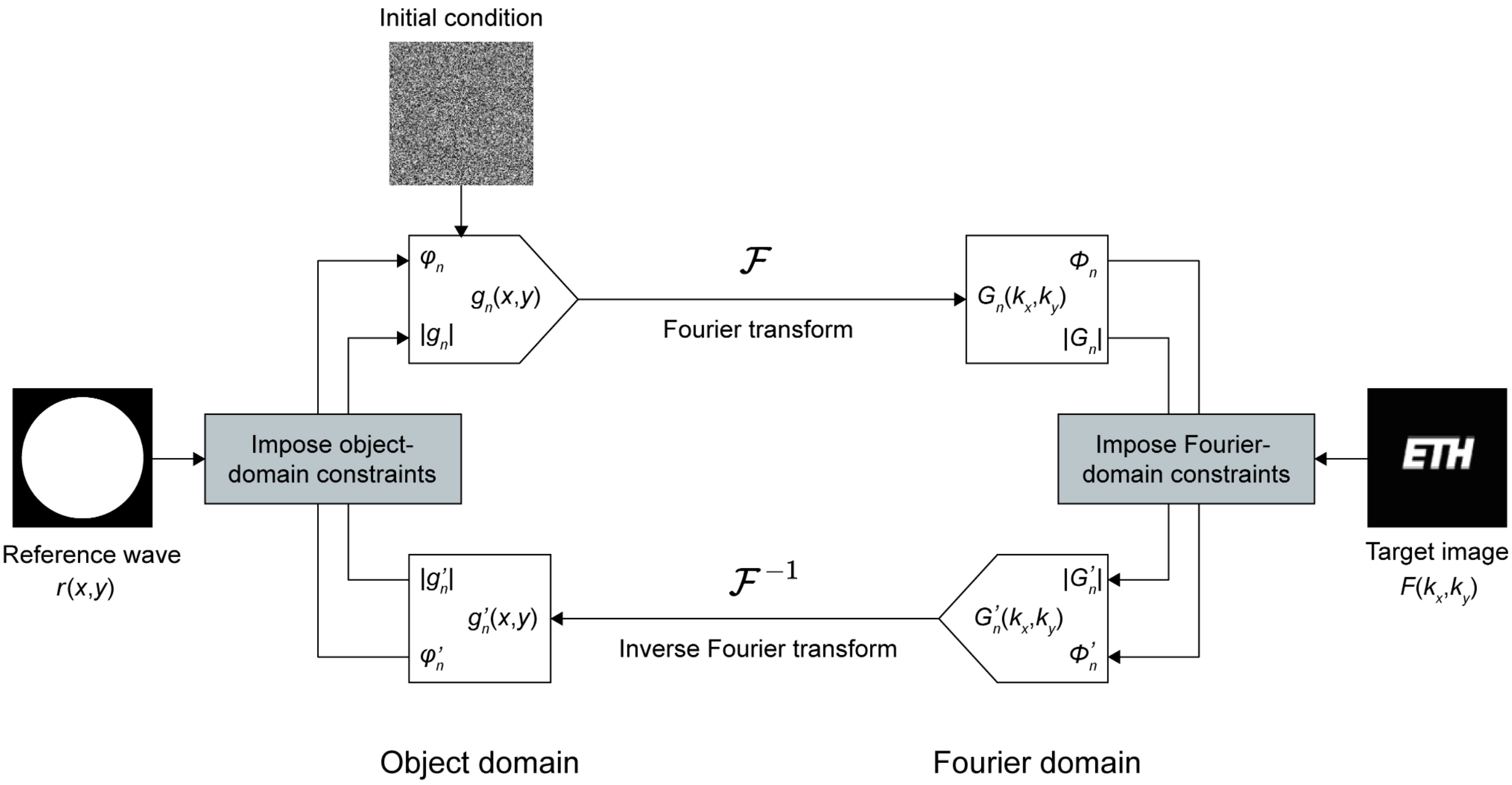

**Figure S11. Schematic of the generalized GS algorithm.** Each iteration of the algorithm consists of four steps (see Section S1.7 for details). Steps 1 and 3 perform the forward and inverse Fourier transform, respectively, to transition between the object and Fourier domains. In both domains, the optical fields are represented by their amplitude and phase. Step 2 imposes the Fourier-domain constraints by applying the target diffraction image $F(k_x, k_y)$. Step 4 enforces the object-domain constraints, which typically include information from the reference wave $r(x, y)$ or fabrication limitations.

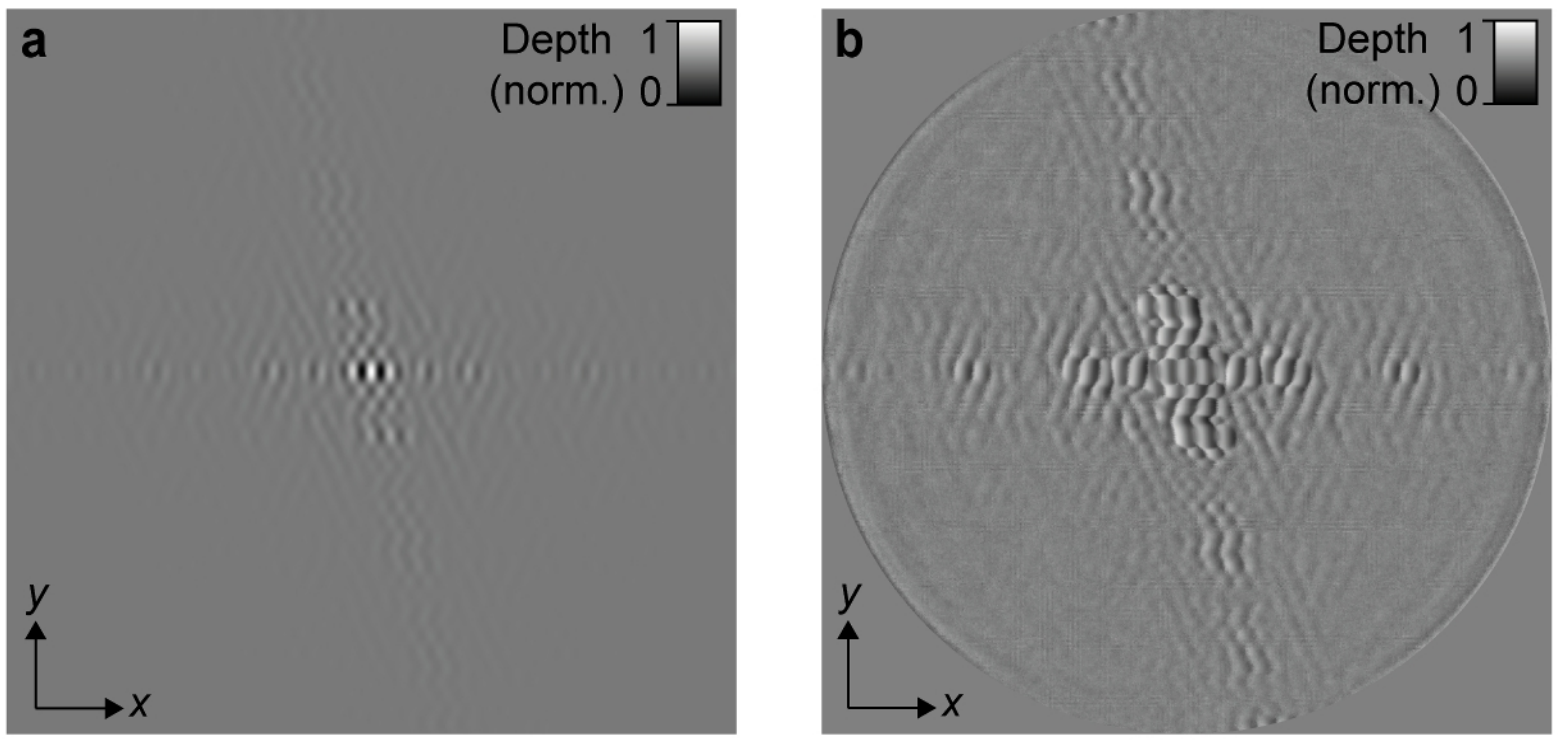

**Figure S12. IFTA with additional phase constraints in the Fourier domain.** (a) Grayscale design generated using the linear design approach for the ETH logo (see Figure 2 in the main text). (b) Corresponding design obtained with a modified GS algorithm that imposes both amplitude and phase constraints in the Fourier domain to suppress speckles in the diffraction image. The algorithm follows the method of Chang *et al.*,[S13] where we set the parameters $(\alpha, \beta, \gamma) = (0.2, 0.5, 1.0)$. The target image mask consisted of a single ETH logo (without its mirror image) with a uniform phase $\varphi = 0$, while a random phase distribution was used as the initial condition.

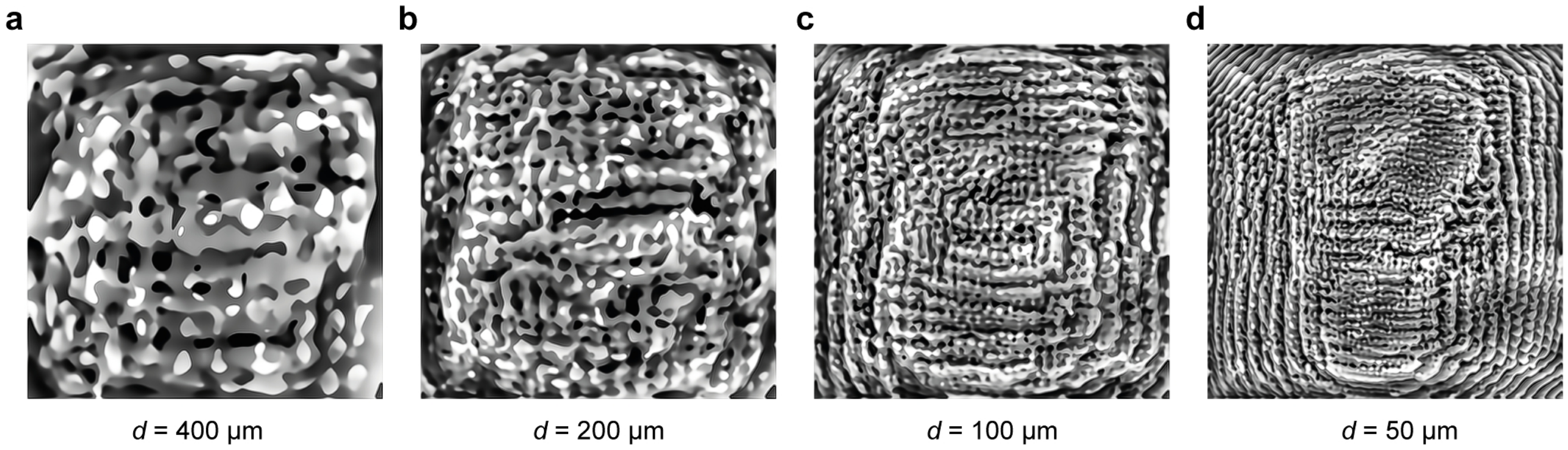

**Figure S13. Feature size as a function of image-plane distance.** (a)–(d) Grayscale designs of wavelength-multiplexed OFS holograms designed to project images to planes located at distances $d$ = 400, 200, 100, and 50 µm away from the sample plane, respectively. The diffraction images at wavelengths 450, 575, and 700 nm ($N_w$ = 3) correspond to handwritten digits 0, 1, and 2 from the Modified National Institute of Standards and Technology (MNIST) dataset (see Figure 4 in the main text). The lateral hologram size was fixed at 50 µm × 50 µm with a pixel size of 40 nm × 40 nm (1250 × 1250 pixels, as in Figure 4). Grayscale depth values range from 0 to 255 (linear scale), representing a total physical depth of 350 nm. We observe that the structural feature size generally grows as distance $d$ increases, consistent with observations based on diffractive lenses with increasing focal distance ($d = f$, see Section S1.3 for details). Intuitively, stronger light bending requires finer structural features.

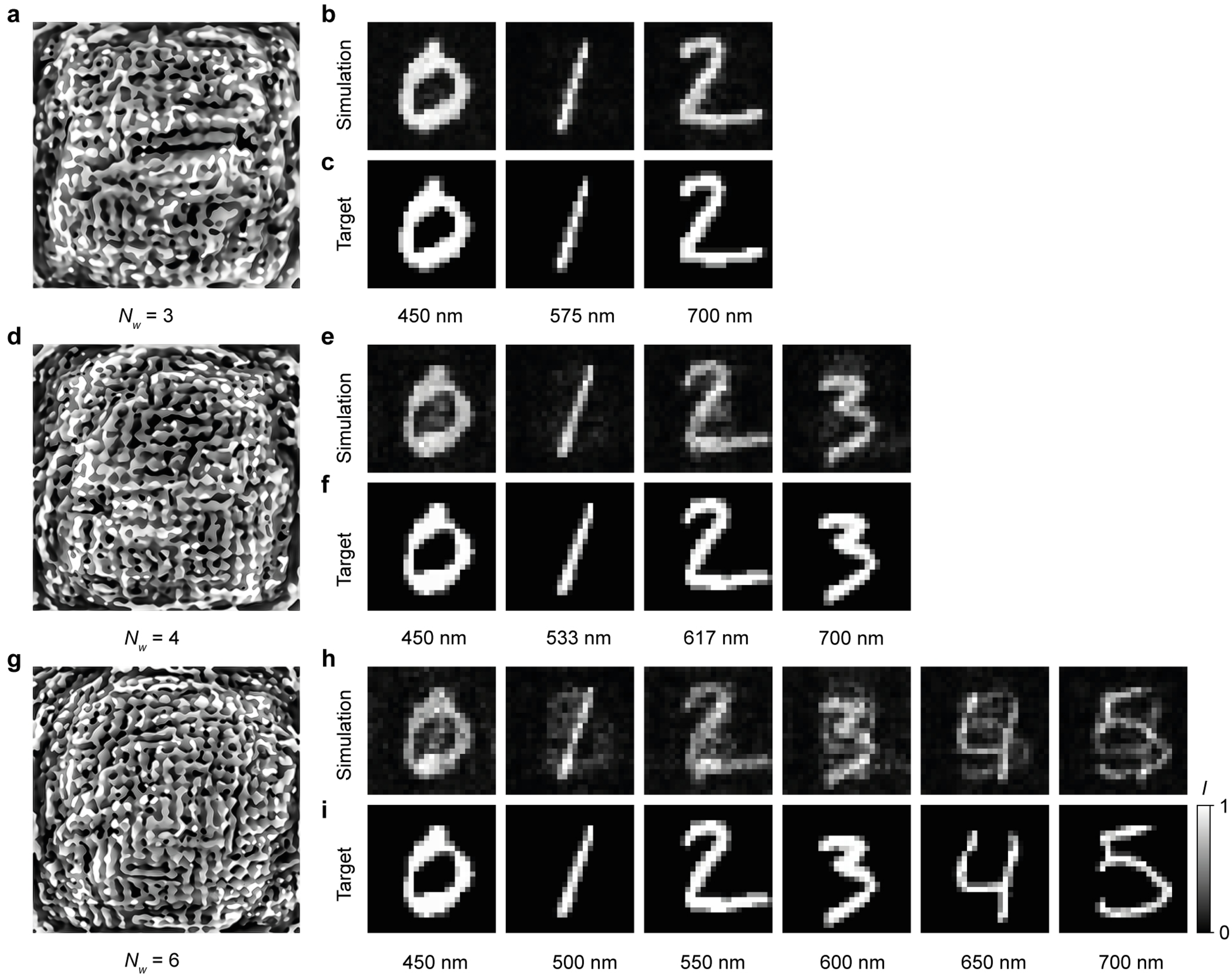

**Figure S14. Impact of the number of wavelength channels $N_w$ on the image reconstruction performance.** (a),(d),(g) Grayscale hologram designs with different numbers of wavelength channels ($N_w$ = 3, 4, and 6). The distance between the sample and image plane was fixed at $d$ = 200 µm. Each hologram measures 50 µm × 50 µm with a pixel size of 40 nm × 40 nm (1250 × 1250 pixels, as in Figure S13 and Figure 4 in the main text). The encoded diffraction images are MNIST handwritten digits from 0 to $N_w - 1$. The wavelength channels are uniformly spaced between 450 to 700 nm. (b),(e),(h) Simulated intensity profiles compared with (c), (f), and (i) the corresponding target images. Intensities are normalized to the maximum value within each panel. As $N_w$ increases, the wavelength channels become spectrally more closely spaced. While diffraction efficiency remains similar (due to the efficiency threshold in the loss function; see Section S1.8), image quality degrades due to enhanced crosstalk between the wavelength channels.

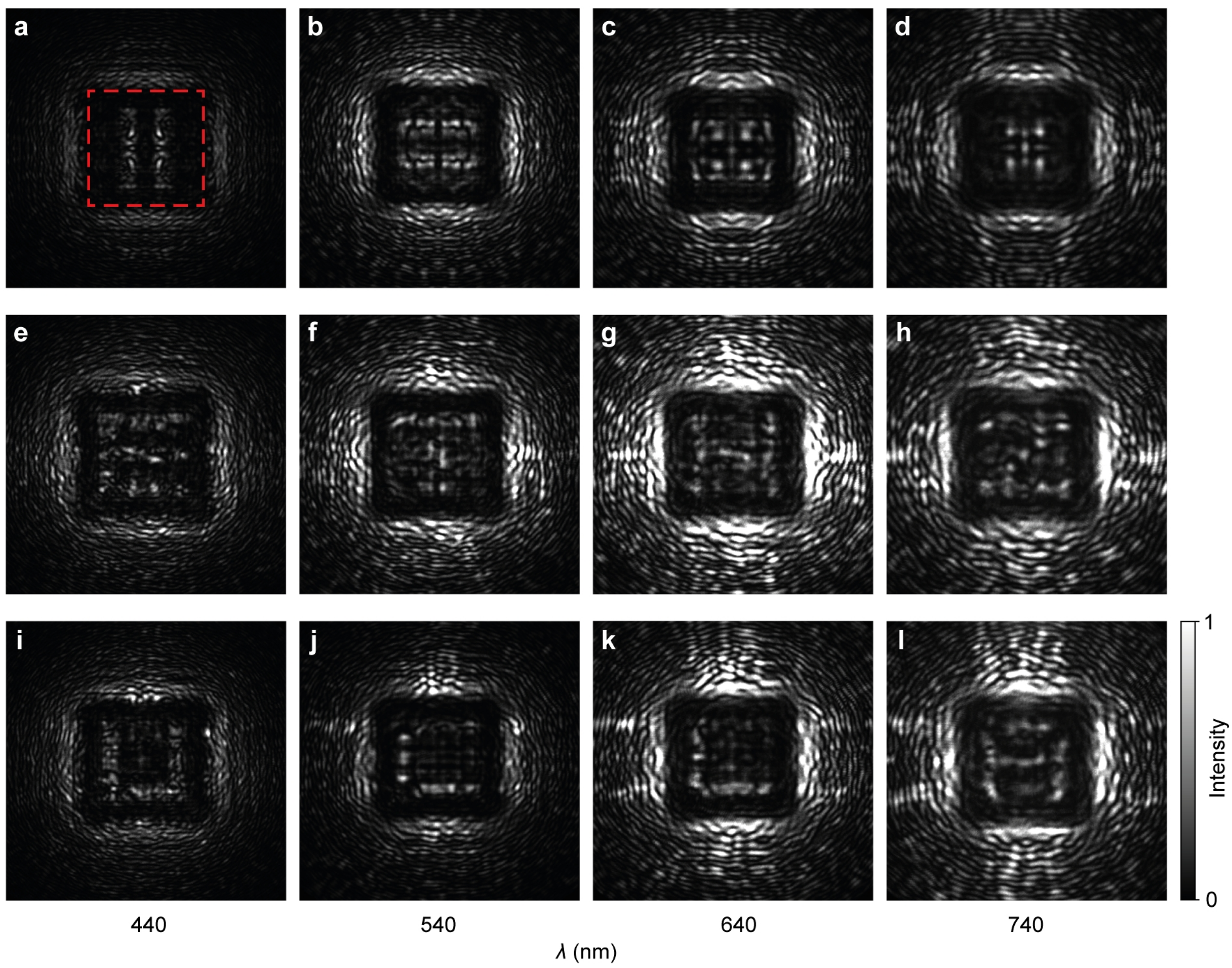

**Figure S15. Raw experimental data for wavelength-multiplexed OFS holograms ($N_w$ = 4).** (a)–(l) Raw experimental intensity patterns for the holograms shown in Figure 5, designed to produce geometric shapes [squares and rectangles; (a)–(d)], the letters "ETHZ" (e)–(h), or the letters "UCLA" (i)–(l), respectively. Intensities are normalized to the maximum value within the field of view (FOV) of each panel (50 μm × 50 μm), indicated by the red dashed square in (a).

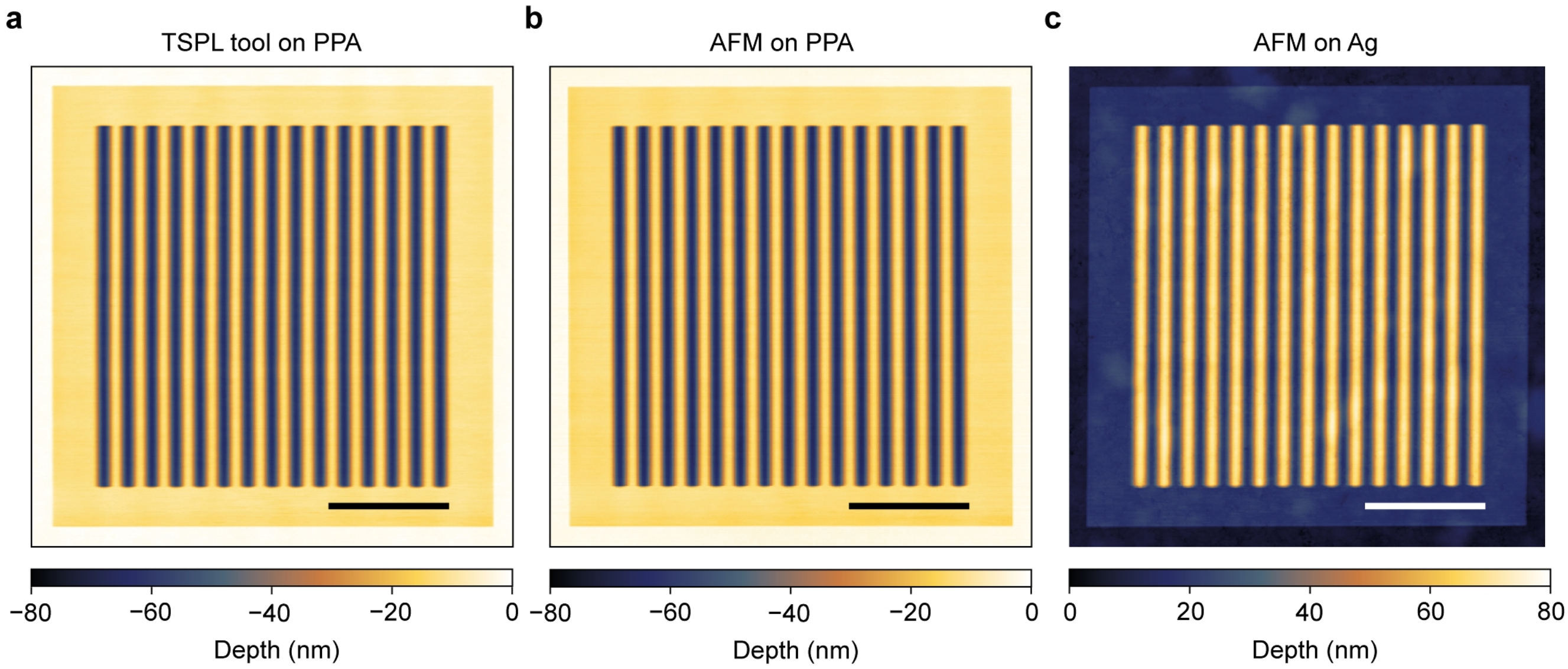

**Figure S16. Topography characterization of a sinusoidal reference structure.** (a) Topography data of the structure in PPA measured by the TSPL tool. (b) Surface profiles in PPA and (c) in Ag measured through atomic-force microscopy (AFM). The reference structure consists of a sinusoid ($\Lambda$ = 600 nm, $A$ = 25 nm) surrounded by a 1-μm-wide patterned frame. Scalebars represent 3 μm.

## S3. SUPPORTING TABLES

**Table S1. Efficiency and peak signal-to-noise (PSNR) of wavelength-multiplexed holograms ($N_w$ = 3).** Complementary data to Figure 4 in the main text.

| MNIST digits | Simulation | | Experiment | |
|---|---|---|---|---|
| Wavelength (nm) | Efficiency (%) | PSNR | Efficiency (%) | PSNR |
| 450 | 8.6 | 20.4 | 8.4 | 10.8 |
| 575 | 4.3 | 22.2 | 8.7 | 17.5 |
| 700 | 6.3 | 20.4 | 10.4 | 13.1 |

**Table S2. Efficiency and PSNR of wavelength-multiplexed holograms ($N_w$ = 4).** Complementary data to Figure 5 in the main text.

| Geometric shapes | Simulation | | Experiment | |
|---|---|---|---|---|
| Wavelength (nm) | Efficiency (%) | PSNR | Efficiency (%) | PSNR |
| 440 | 11.8 | 20.1 | 10.4 | 10.2 |
| 540 | 7.4 | 20.2 | 12.7 | 11.0 |
| 640 | 6.8 | 19.5 | 9.2 | 9.9 |
| 740 | 7.6 | 19.9 | 7.8 | 10.2 |

| ETHZ | Simulation | | Experiment | |
|---|---|---|---|---|
| Wavelength (nm) | Efficiency (%) | PSNR | Efficiency (%) | PSNR |
| 440 | 9.3 | 18.9 | 10.0 | 8.7 |
| 540 | 4.6 | 19.3 | 9.2 | 10.2 |
| 640 | 3.5 | 18.8 | 6.2 | 8.8 |
| 740 | 6.4 | 18.9 | 7.5 | 8.9 |

| UCLA | Simulation | | Experiment | |
|---|---|---|---|---|
| Wavelength (nm) | Efficiency (%) | PSNR | Efficiency (%) | PSNR |
| 440 | 10.6 | 20.3 | 10.2 | 9.3 |
| 540 | 5.7 | 20.6 | 10.8 | 10.9 |
| 640 | 3.7 | 20.2 | 7.1 | 10.7 |
| 740 | 5.6 | 20.4 | 7.5 | 8.7 |

**Table S3. Topography characterization of a sinusoidal reference structure.** The uncertainties of the fitted periods $\Lambda$ and amplitudes $A$ are below 0.01 nm.

| | TSPL tool on PPA | AFM on PPA | AFM on Ag |
|---|---|---|---|
| Fit parameter | Value (nm) | Value (nm) | Value (nm) |
| $\Lambda$ | 601 | 605 | 604 |
| $A$ | 24.9 | 26.0 | 27.3 |
| RMSE | 0.94 | 1.46 | 1.70 |